\documentclass[twocolumn]{IEEEtran} 
\usepackage[T1]{fontenc}
\usepackage{color}
\usepackage{mathrsfs}
\usepackage{amsmath}
\usepackage{amsthm} 
\usepackage{amssymb}
\usepackage{graphicx}
\usepackage{cite}
\usepackage{makecell}
\usepackage{algorithm}
\usepackage{algorithmic}
\usepackage{fancyhdr}
\usepackage{diagbox}
\usepackage{bm}
\usepackage[unicode=true,
 bookmarks=true,bookmarksnumbered=true,bookmarksopen=true,bookmarksopenlevel=1,
 breaklinks=false,pdfborder={0 0 0},pdfborderstyle={},backref=false,colorlinks=false]
 {hyperref}
\hypersetup{
 pdftitle={Your Title},
 pdfauthor={Your Name},
 pdfpagelayout=OneColumn, pdfnewwindow=true, pdfstartview=XYZ, plainpages=false}

\usepackage{multirow} 
\usepackage{xcolor}

\allowdisplaybreaks[4]

\usepackage[caption=false,font=footnotesize]{subfig}

\IEEEoverridecommandlockouts

\usepackage{geometry}
\begin{document}
\bibliographystyle{IEEEtran}

\title{\textcolor{black}{Cooperative UAV and Ground Station for Robust MEC of Dynamic Aquatic IoT: A Generative AI-Enhanced Approach}

\author{\IEEEauthorblockN{Jiahao You, \IEEEmembership{Graduate Student Member,~IEEE,} Ziye Jia, \textit{Member, IEEE,} Chao Dong, \IEEEmembership{Senior Member,~IEEE,} \\Qihui Wu, \IEEEmembership{Fellow,~IEEE,} and Zhu Han, \IEEEmembership{Fellow,~IEEE}}
\thanks{Jiahao You, Ziye Jia, Chao Dong, and Qihui Wu are with the Key Laboratory of Dynamic Cognitive System of Electromagnetic Spectrum Space, Nanjing University of Aeronautics and Astronautics, Nanjing 211106, China (e-mail: yjiahao@nuaa.edu.cn, jiaziye@nuaa.edu.cn, dch@nuaa.edu.cn, wuqihui@nuaa.edu.cn). 
\textit{(Corresponding authors: Ziye Jia and Chao Dong.)}

Zhu Han is with the Department of Electrical and Computer Engineering at the University of Houston, Houston, TX 77004 USA (e-mail: hanzhu22@gmail.com).}}}

\maketitle
\begin{abstract}
The increasing aquatic Internet of Things (AIoTs) require computational support and coverage in applications such as maritime search and rescue. 
Unmanned aerial vehicles (UAVs) can offer low-cost, flexible aerial services, and ground stations (GSs) can provide powerful supports, which can cooperate to help the AIoTs in complex scenarios.
However, the collaboration between UAVs and GSs for AIoTs faces challenges of task uncertainties, AIoTs trajectory uncertainties, heterogeneities, and limited computational resources.
To address these issues, we propose a cooperative UAV and GS based robust multi-access edge computing (MEC) framework to assist AIoTs in completing computational tasks. 
Specifically, we formulate the optimization problem of joint task offloading and UAV trajectory to minimize the total execution time, 
which is in the form of mixed integer nonlinear programming and NP-hard to tackle.
Therefore, we propose the algorithm of generative artificial intelligence-enhanced heterogeneous agent proximal policy optimization (GAI-HAPPO). 
The proposed algorithm integrates GAI models to enhance the actor network ability to model complex environments and extract high-level features, thereby allowing the algorithm to predict uncertainties and adapt to dynamic conditions. 
Additionally, GAI stabilizes the critic network, addressing the instability of multi-agent reinforcement learning approaches.
Finally, extensive simulations demonstrate that the proposed algorithm outperforms the existing benchmark methods, thus highlighting the potentials in tackling intricate, cross-domain issues in the considered scenarios.
\end{abstract}


\section{Introduction}\label{s1}
\IEEEPARstart{I}{n} recent years, the demand for efficient, responsive, and scalable solutions has significantly increased in fields such as environmental monitoring, maritime search and rescue, where operations often span both air and water environments \cite{Survey_iot_2022, Energy_tvt_2024, JCOR_2025}. 
These tasks rely on isolated systems, with aquatic Internet of Things (AIoTs) handling water-based operations and unmanned aerial vehicles (UAVs) or ground stations (GSs) operating separately in respective domains \cite{Distributionally_2025_cc, Service_cyl_2025}. 
However, constrained by limited onboard energy, computing capacity, and communication range, AIoTs are unable to meet the intensive demands of real-time data processing and wide-area task coverage.
Given these constraints, AIoTs alone are ill-suited for real-time coordination and processing. UAVs serve as agile relays that extend communication coverage, while GSs offer reliable, high-capacity computation. Therefore, they complement the limitations of AIoTs in both connectivity and processing.

Building on this complementarity, we propose a cooperative system where AIoTs generate and relay tasks, while UAVs and GSs provide flexible coverage and enhanced computation to collaboratively execute them, thereby improving overall efficiency and responsiveness \cite{Integrating_2024, Opt_DSS_2025, Cooperative_2024, Luo_DRL_TP_2023}.
To manage these distributed resources effectively, the multi-access edge computing (MEC) is introduced to facilitate task offloading and optimize resource allocation across AIoTs, UAVs, and GSs \cite{Resource_2023, Distributed_DSS_2025}. 
However, the collaboration among AIoTs, UAVs, and GSs in an MEC environment must contend with inherent challenges arising from real-world dynamics and system heterogeneity.
The unpredictable movement patterns of AIoTs and the variability in task arrival times pose significant challenges for robust coordination, requiring UAVs and GSs to adapt resiliently to these dynamic changes \cite{UAV_2024, Mobility_2015, Online_2022}.
Additionally, the heterogeneous of these devices, with each equipped with different communication and processing capabilities, aggravates the complexity to effective resource sharing and coordination \cite{Collaborative_2024, HEE_2023, Learning_Min_2019}. 

The combination of random AIoT positions, task uncertainty, and device heterogeneity bring more challenges for effective MEC implementation.
Specifically, the device heterogeneity complicates unified task allocation, task uncertainty causes unpredictable computational loads that challenge queue stability, and random AIoT positions lead to frequently changing link conditions that hinder consistent offloading and real-time coordination.
To tackle these issues, the multi-agent reinforcement learning (MARL) methods can be employed to deal with the cooperative MEC \cite{Beyond_2024, Massive_2023,Learning_Privacy_Min_2019}. Through interactions with the environment, agents continuously learn by trial and error, enhancing their dynamic perception and predicting uncertainties. 
However, current MARL algorithms still have limitations, for  instance, the actor network struggles to model complex environments and is unable to extract high-level features from them effectively \cite{Approximation_2022, GAI_2024}. 
Moreover, due to inherent uncertainties in agent-environment interactions, the critic network training often becomes unstable.
Besides, the generative artificial intelligence (GAI) offers a promising solution to these issues \cite{GAI_2024}.
Since GAI is capable of learning data distributions, capturing underlying structures, which can be widely applied in fields such as content analysis and creation, including ChatGPT and Sora \cite{GPT_2023}. With higher sampling efficiency, improved generalization, and robustness to environmental variations, GAI can be leveraged to improve MARL algorithms.

In light of these challenges and the potential of GAI, we investigate the collaboration problem among AIoTs, UAVs, and GSs in this paper, considering the random AIoT position and the uncertain task arrival. Specifically, the AIoT movement is assumed following the Gauss-Markov mobility model \cite{Mobility_2015,jia2025,Online_2022}. Additionally, the task arrival follows the Poisson distribution in a collaborative framework with heterogeneous devices \cite{Enhance_2020, NFV_cyl_2025}. 
Then, we propose a cooperative UAV and GS based robust MEC framework to assist AIoTs in completing computational tasks.
In detail, we formulate the optimization problem of joint task offloading and UAV trajectory to minimize the total execution time, 
which is in the form of mixed integer nonlinear programming (MINLP) and NP-hard to tackle \cite{Joint_Chu_2024,Joint_Hoang_2024}.
Hence, we propose GAI-enhanced heterogeneous agent proximal policy optimization (GAI-HAPPO) and utilize GAI to enhance the algorithm performance. 
To enhance the adaptability and expressiveness of multi-agent policy learning, we incorporate two key components inspired by GAI: a transformer-based actor network and a generative adversarial network (GAN)-based critic network. The transformer enables the actor to model spatial-temporal dependencies and heterogeneous agent interactions, while the GAN refines the action-value estimation through adversarial learning. These components together enhance decision quality and training robustness, forming the foundation of our GAI-enhanced framework.

In summary, our contributions are summarized as follows.
\begin{itemize}
\item We propose a robust cooperative MEC framework integrating UAVs and GSs to effectively assist AIoTs in computational task execution. Different from existing works that typically handle UAVs and GSs independently, our framework jointly exploits the flexible mobility and aerial coverage of UAVs and the superior computational resources of GSs, significantly enhancing the real-time computational support capability in dynamic aquatic environments.
\item We formulate a joint optimization problem to simultaneously determine task offloading decisions and UAV trajectories under heterogeneous MEC conditions. The formulated model explicitly captures uncertainties stemming from random AIoT positions and stochastic task arrivals, aiming to minimize the total task execution time. The resulting formulation is characterized as a MINLP problem, which is NP-hard and challenging to solve directly.
\item We develop a algorithm termed as GAI-HAPPO, to efficiently solve the formulated MINLP. Specifically, by integrating GAI components, including a transformer-based actor network and a GAN-based critic network, GAI-HAPPO significantly improves feature extraction and stabilizes value estimation, thereby enhancing the adaptability and training stability in complex, uncertain environments.
\item We conduct extensive numerical simulations to validate the effectiveness and superiority of the proposed approach. Simulation results demonstrate that GAI-HAPPO achieves a performance improvement of 22.8\% over the original HAPPO baseline, showing its advantage in handling dynamic, heterogeneous, and cross-domain challenges in cooperative UAV and GS-based MEC scenarios.
\end{itemize}

The rest of the paper is organized as follows.
Section \ref{s2} discusses related works.
Section \ref{s3} details the system model and problem formulation. 
Section \ref{s4} introduces the GAI-HAPPO algorithm. 
Section \ref{s5} presents numerical results, and finally conclusions are drawn in Section \ref{s6}.

\section{Related Work}\label{s2}
\subsection{Collaboration in AIoT-UAV Systems}
The integration of AIoTs with unmanned aerial and surface platforms receives increasing attentions, as it enables wide-area perception, enhanced communication, and task coordination across air-water environments. In real-world applications, \cite{USV_uav_2019} presents a USV-UAV platform designed for collaborative missions, which includes adjustable buoys, a dedicated carrier deck, integrated sensors, and a multi-ultrasonic positioning system. Building on these studies, \cite{Energy_tvt_2024, Integrating_2024, 3u_2023} design cooperative task schemes involving AIoTs and UAVs. For instance, in maritime search and rescue scenarios, \cite{Integrating_2024} proposes a robust adaptive event-triggered control strategy for a USV-UAV cooperative system, aiming to handle sensor faults and reduce communication overhead. \cite{3u_2023} incorporates underwater vehicles and constructs a network of UAVs, USVs, and unmanned underwater vehicles to perform underwater target hunting collaboratively. To address communication challenges in inland waterways, \cite{Energy_tvt_2024} enhances wireless performance by integrating an intelligent reflecting surface into a hybrid USV-UAV network. These studies validate the feasibility of AIoT-UAV/USV cooperation, but overlook task offloading and coordination challenges arising from MEC heterogeneity.

\begin{figure*}[htbp]
\centering
\includegraphics[width=14cm]{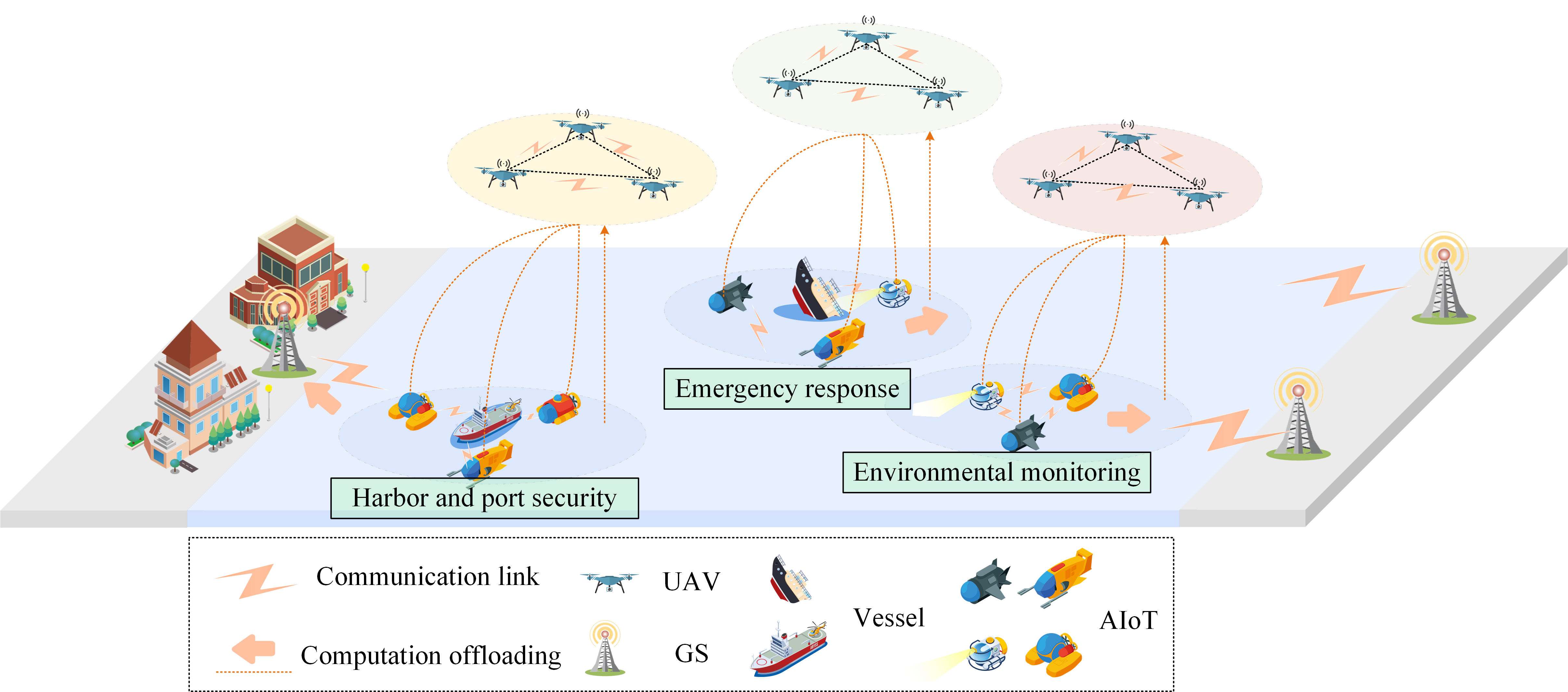}
\caption{System model overview.}
\label{f1}
\end{figure*}

\subsection{Uncertainty in MEC}
The uncertainty in MEC environments, such as random task arrivals, user mobility, and dynamic resource availability, poses significant challenges to efficient and reliable system performance. Recent works address these issues through optimization-based and adaptive task management strategies.
For example, \cite{Distributed_Xia_2024} develops a distributed delay-constrained computation offloading scheme for ultra-dense MEC networks by combining game theory with Lyapunov optimization, enabling adaptive task handling under user mobility and limited resources. \cite{Uncertainty_Apostolopoulos_2023} proposes a data offloading framework that optimizes resource utilization in uncertain computing environments. In \cite{Energy_Xu_2022}, an optimization framework addresses uncertain payoffs in collaborative service caching and request offloading, aiming to balance the efficiency under variable rewards. \cite{Joint_ren_2020} presents a joint channel allocation and resource management strategy for hierarchical MEC systems, using Lyapunov optimization to address stochastic arrivals and fluctuating capacity.
However, these studies primarily adopt model-based approaches and lack adaptive mechanisms to maintain MARL stability under unpredictable task arrivals and mobility patterns in MEC.

\subsection{Heterogeneity in MEC}
Heterogeneous device capabilities and network structures pose significant challenges for efficient computation offloading and resource allocation in MEC systems. \cite{Computation_zhang_2023} models a spatially random two-tier heterogeneous network with macro and small-cell base stations of varying computing power, and designs offloading strategies based on the stochastic geometry and queuing theory. \cite{Heterogeneous_ke_2020} proposes an adaptive offloading framework for heterogeneous vehicular networks, using deep reinforcement learning (DRL) to jointly optimize the energy and delay under stochastic wireless conditions. In addition, \cite{HetMEC_Zhang_2020} introduces a heterogeneous multi-layer MEC framework for 6G, where RL is used to coordinate task offloading, spectrum access, and control across layers of diverse devices. Although these studies consider heterogeneous MEC environments, they often assume static configurations and lack coordination strategies suitable for diverse agents operating under dynamic conditions.

\subsection{GAI-Enhanced RL}
GAI, supported by large models such as ChatGPT, increasingly contributes to RL by improving the generalization, robustness, and policy modeling.
\cite{GAI_UAV_2024} explores the use of GAI in UAV systems, incorporating GANs, variational autoencoders, diffusion models, and transformers. \cite{GAI_2024} applies GAI to improve DRL from both data augmentation and policy optimization perspectives. \cite{gan_fl_2024} proposes a GAN-based federated learning framework, termed collaborative game parallel learning, which addresses non-IID data challenges and enables personalized model training at the network edge via a mix-generator module. In large-scale MEC systems, \cite{gan_mec_2024} investigates delay-optimal task offloading using a mean-field game formulation and GAN-based estimation of unknown task distributions. \cite{Diffusion_Reinforcement_2024} combines diffusion-generated decisions with RL to improve the decision-making under uncertainties in artificial intelligence-generated content services. Despite the promise of GAI in improving learning performance, most existing works do not address the stability issues of MARL in uncertain environments or the coordination challenges caused by heterogeneous multi-agent MEC systems.

In summary, GAI serves as a powerful enhancement to RL by improving generalization, sample efficiency, and robustness to uncertainty. These properties are especially valuable in multi-agent MEC systems, where agents operate under partial observability, dynamic task arrivals, and heterogeneous constraints. Motivated by these strengths, our work proposes GAI-HAPPO, which integrates a transformer-based actor and a GAN-based critic to enhance policy modeling and value estimation for cooperative AIoT applications.

\begin{table*}[h]
\caption{Key Notations}
\begin{center}
\begin{tabular}{|m{3.8cm}<{\centering}|m{11cm}|}
\hline
\textbf{Symbol} &\textbf{Description}\\
\hline
$\mathbb{S}, \mathbb{U}, \mathbb{G}$&Set of AIoTs, UAVs, and GSs, respectively.\\
\hline
$i,j,k$&Index of AIoTs $\mathbb{S}$, UAVs $\mathbb{U}$, and GSs $\mathbb{G}$, respectively.\\
\hline
$\mathbb{T}, \tau$&Set of time slots $\mathbb{T}$ and the length of each time slot $\tau$, respectively.\\
\hline
$A_i(t), d_i(t), c_i(t)$ &Task generated by AIoT $i$ at time slot $t$, where $d_i(t)$ is the data size, and $c_i(t)$ is the required computation in CPU cycles per bit.\\
\hline
$\lambda_i(t)$ &Mean task arrival rate for AIoT $i$, modeled by a Poisson distribution.\\
\hline
$l_i^s(t), l^u_j(t), l^g_k(t)$ &Position of AIoT $i$, UAV $j$, and GS $k$ at time slot $t$, respectively.\\
\hline
$\xi_{i,j}(t), R_{i,j}^{a2u}(t)$&Path loss and transmission rate between AIoT $i$ and UAV $j$ at time slot $t$, respectively.\\
\hline
$G_{i,k}(t), R_{i,k}^{a2g}(t)$&Channel power gain and transmission rate from AIoT $i$ to GS $k$ at time slot $t$, respectively.\\
\hline
$v_i(t)$&Velocity of AIoT $i$ at time slot $t$.\\
\hline
$\theta_j(t), k_j(t)$&Horizontal azimuth $\theta_j(t)$ and flying distance $k_j(t)$ of UAV $j$ at time slot $t$, respectively.\\
\hline
$p_{i,j}(t), q_{i,k}(t)$&Offloading decision variables from AIoT $i$ to UAV $j$, and from AIoT $i$ to GS $k$ at time slot $t$, respectively.\\
\hline
$\alpha_i(t), \beta_{i}(t), \gamma_{i}(t)$&Workload fractions processed by AIoT $i$, offloaded to UAV $j$, and offloaded to GS $k$ at time slot $t$, respectively.\\
\hline
$f_i^s, f_j^u, f_k^g$&Computation resources of AIoT $i$, UAV $j$, and GS $k$, respectively.\\
\hline
$Q_i^s(t), Q_j^u(t), Q_k^g(t)$&Task queue sizes of AIoT $i$, UAV $j$, and GS $k$ at time slot $t$, respectively.\\
\hline
$T_{i}^{s}(t), T_{i,j}^{u}(t), T_{i,k}^{g}(t)$&Computation time for task $A_i(t)$ processed by AIoT $i$, UAV $j$, and GS $k$, respectively.\\
\hline
$\Phi(t)$&Total time cost for all tasks generated by AIoTs at time slot $t$.\\
\hline
\end{tabular}\label{T1}
\end{center}
\end{table*}

\section{System Model}\label{s3}
As illustrated in Fig. \ref{f1}, the considered system model consists of AIoTs, UAVs, and GSs. UAVs are equipped with MEC servers to provide support to AIoTs, while GSs possess powerful computing capabilities. 
Specifically, the set of AIoTs is represented by $\mathbb{S} = \{1,2,...,I\}$, the set of UAVs by $\mathbb{U} = \{1,2,...,J\}$, and the set of GSs by $\mathbb{G} =\{1,2,...,K\}$. The considered time duration is divided into multiple time slots, denoted as $\mathbb{T} = \{1, 2,..., T\}$, each with length of $\tau$.
In time slot $t$, the task generated by AIoT $i$ is denoted as $A_i(t) = \{d_i(t), c_i(t)\}$, where $d_i(t)$ refers to the data size in bit, and $c_i(t)$ represents the required computation resources in CPU cycles per bit. To better simulate random events occurring per unit time in real scenarios, we utilize the Poisson distribution with the mean value of $\mathbb{E}(A_i(t)) = \lambda_i$ to model the generation of AIoT workloads. The corresponding random variable is represented as $A(t)=\{A_1(t),...,A_I(t)\}$, where $A_i(t)$ represents the total workload arriving for AIoT $i$ during time slot $t$.

AIoTs carry out tasks such as environmental monitoring and emergency rescue, which are divided into three main stages: 
\begin{itemize}
    \item Stage \uppercase\expandafter{\romannumeral1} involves the local computation time of the AIoTs.
    \item Stage \uppercase\expandafter{\romannumeral2} includes both the transmission time and the computation time of the UAVs.
    \item Stage \uppercase\expandafter{\romannumeral3} covers the transmission and computation times of the GS.
\end{itemize}
AIoTs have the option to offload part of their workload to UAVs. However, due to the resource limitations of UAVs, it can result in high computational latency. On the other hand, although GSs have powerful computational abilities that can reduce computational latency, they encounter substantial communication latency.

\subsection{Communication Model}
The positions of AIoTs, UAVs, and GSs are depicted by the three-dimensional cartesian coordinate, i.e., $l_i^s(t) = (x_i^{s}(t), y_i^{s}(t), h_i^s(t))$, $l^u_j(t) = (x_j^{u}(t), y_j^{u}(t),h_j^u(t))$, and $l^g_k(t) = (x_k^{g}(t), y_k^{g}(t), h_k^g(t))$, respectively. 
Since the network remains quasi-static within each time slot, the channel state within a given time slot can be considered quasi-static. When the time slot changes, the channel state can be accurately and promptly updated based on the following channel model and new parameters.

\subsubsection{AIoT to UAV (A2U) Channel Model}
The path loss from AIoT $i$ to UAV $j$ is 
\begin{equation}
\begin{split}
\xi_{i,j}(t) &= \frac{\zeta_{L}-\zeta_{NL}}{1+a \exp \{-b [\varsigma_{i,j}(t) -a \}} \\
& + 20 \lg\left(\frac{4\pi f_c\|l_i^s(t)-l_j^u(t)\| }{C}\right)+\zeta_{NL},
\end{split}
\end{equation}
where
\begin{equation}
\begin{aligned}
\varsigma_{i,j}(t) = \arctan \left(\frac{\vert h_i^s-h_j^u \vert }{\|l_i^s(t)-l_j^u(t)\|_2 }\right),
\end{aligned}
\end{equation}
in which $f_c$ denotes the carrier frequency, and $C$ represents the speed of light.
$\zeta_{L}$, $\zeta_{NL}$, $a$, and $b$ are environment related  parameters\cite{wang_hybrid_2021, Jia_Joint_HAP_2021}.
According to the Shannon formula, the average rate between AIoT $i$ and UAV $j$ is computed as 
\begin{equation}
R_{i,j}^{a2u}(t)= B_{i,j}^{a2u}(t) \log_2  \left(1+\frac{P_{i}^s\xi_{i,j}(t)}{N_G}\right),
\end{equation}
where $B_{i,j}^{a2u}(t)$ is the available bandwidth at UAV $j$ from AIoT $i$, $P_{i}^s$ represents the transmitting power of AIoT $i$, and $N_G$ indicates the power of the additive white Gaussian noise.

\subsubsection{AIoT to GS (A2G) Channel Model}
The time-varying channel power gain from AIoT $i$ to GS $k$ in time slot $t$ is expressed as
\begin{equation}
G_{i,k}(t) = G_0(\|l_i^s(t)-l_k^g(t)\|_2)^{-2},
\end{equation}
where $G_0$ is the channel power gain when $\|l_i^s(t)-l_k^g(t)\|_2=1 \text{m}$ \cite{Luo_DRL_TP_2023, Jia-Hierarchical_2022}. 
Consequently, the transmission rate from AIoT $i$ to GS $k$ is 
\begin{equation}
R_{i,k}^{a2g} = B_{i,k}^{a2g}(t) log_2\left(1+\frac{P_{i}^sG_{i,k}(t)}{N_G}\right),
\end{equation}
where $B_{i,k}^{a2g}(t)$ represents the available bandwidth at GS $k$ from AIoT $i$.

\subsection{AIoT Mobility Model}
It is assumed that AIoTs follow the Gauss-Markov mobility model \cite{Mobility_2015,Online_2022}.
Specifically, the velocity of AIoT $i$ at time slot $t$ can be described as
\begin{equation}
v_i(t+1) = \mu  v_i(t) + (1 - \mu) \overline{v} + \overline{\sigma} \sqrt{1-\mu^2}w_i(t),
\end{equation}
where $v_i(t) = (v_i^x(t),v_i^y(t), 0)$ denotes the velocity vector, and $w_i(t)=(w_i^x(t),w_i^y(t), 0)$ represents an uncorrelated random Gaussian process $\mathcal{N}(0,\sigma^2)$. $\mu$, $\overline{v}$, and $\overline{\sigma}$ denote the memory level, asymptotic mean, and asymptotic standard deviation of velocity, respectively. Therefore, the location is updated as
\begin{equation}
l^s_i(t+1) = l^s_i(t) + v_i(t)\tau.
\end{equation}
At the beginning of time slot $t$, it is assumed that each UAV knows the current locations $\{l^s_i(t)\}_{i=1}^I$ of all $I$ AIoTs based on the location feedback from the AIoTs. However, the future trajectories of the AIoTs remain unknown to the UAVs.

\subsection{UAV Mobility Model}
To improve the energy efficiency, UAVs are assumed to operate at a fixed altitude \( H \) above the ground. This modeling choice avoids frequent vertical maneuvers such as ascending and descending, which are known to incur significantly higher energy consumption due to additional thrust requirements and aerodynamic drag \cite{Joint_Optimization_Zhao_2025,Tradeoff_Between_Zhan_2024,UAV_Enabled_Xu_2018}. By constraining the UAV mobility to the horizontal plane, the system reduces propulsion energy overhead and enables more tractable optimization of trajectories and offloading strategies within energy-constrained settings.
Consequently, we consider the horizontal flight of UAVs over the target area, which requires modeling the dynamic trajectory of UAVs.
In time slot $t$, the flying horizontal azimuth $\theta_j(t)$ of UAV $j$ and the flying distance $k_j(t)$ should satisfy the following constraints \cite{Joint_Trajectory_Chen_2025,Task_Offloading_Zhang_2024},
\begin{equation}
\theta_j(t) \in [0, 2\pi], \forall j \in \mathbb{U}, \forall t \in \mathbb{T},\label{e6}
\end{equation}
and
\begin{equation}
k_j(t) \in [0, k^{max}], \forall j \in \mathbb{U}, \forall t \in \mathbb{T},\label{e7}
\end{equation}
where $k^{max}$ represents the maximum distance a UAV can fly in time slot $t$, and the flying distance is calculated as
\begin{equation}
k_j(t) = \| l^u_j(t+1) - l^u_j(t)\|.
\end{equation}

Additionally, the horizontal location coordinates $x$ and $y$ of UAV $j$ at time $t$ can be determined using the horizontal azimuth $\theta_j(t)$ and the flying distance $k_j(t)$ as 
\begin{equation}
x_j^{u}(t) = x_j^{u}(0) + \sum_{t=1}^{t} k_j(t) \cos(\theta_j(t)),
\end{equation}
and 
\begin{equation}
y_j^{u}(t) = y_j^{u}(0) + \sum_{t=1}^{t} k_j(t) \sin(\theta_j(t)).
\end{equation}

To ensure UAVs stay within the target area, which affect trajectory decisions, the following constraints are
\begin{equation}
0 \leq x_j^{u}(t) \leq x_{max}, \forall j \in \mathbb{U}, t \in \mathbb{T},\label{e13}
\end{equation}
and
\begin{equation}
0 \leq y_j^{u}(t) \leq y_{max}, \forall j \in \mathbb{U}, t \in \mathbb{T},\label{e14}
\end{equation}
where $x_{max}$ and $y_{max}$ denote the maximum X and Y coordinate values of the target area, respectively.

\subsection{Computation Model}
In each time slot $t$, task $A_i(t)$ generated by AIoT $i$ can be divided into three parts, which are computed by AIoT $i$, UAV $j$, and GS $k$, respectively.
Therefore, we define the offloading variable of AIoT $i$ offloaded to UAV $j$ as $p_{i,j}(t)$: 
\begin{equation}
p_{i,j}(t)= 
\begin{cases}
1, & \text {if AIoT } i \text { offloaded to UAV } j, \\ 
0, & \text {otherwise, }\label{e15}
\end{cases}
\end{equation}
and the offloading variable of AIoT $i$ offloaded to GS $k$ as $q_{i,k}(t)$: 

\begin{equation}
q_{i,k}(t) = 
\begin{cases}
1, & \text {if AIoT } i \text { offloaded to GS } k, \\ 
0, & \text {otherwise. }\label{e16}
\end{cases}
\end{equation}

It is worth noting that for task $A_i(t)$, only one UAV and one GS can be selected for offloading computation within the same time slot, i.e.,
\begin{equation}
\sum_{j=1}^J p_{i,j}(t) \leq 1, j\in\mathbb{U},\label{e17}
\end{equation}
and
\begin{equation}
\sum_{k=1}^K q_{i,k}(t) \leq 1, \forall i \in \mathbb{S}.\label{e18}
\end{equation}

Let $\alpha_i(t)$ $(\alpha_i(t) \in [0,1],\forall i\in\mathbb{S})$, $\beta_i(t)$ $ (\beta_i(t) \in [0,1],\forall i\in\mathbb{S})$, and $\gamma_i(t)$ $(\gamma_i(t) \in [0,1],\forall i\in\mathbb{S})$ denote the proportion of the workload of AIoT $i$ that is computed by AIoT $i$, UAV $j$, and GS $k$, respectively.
In detail, for the task $A_i(t)$ of AIoT $i$, a portion of the workload can be allocated to UAV $j$ and GS $k$.
Therefore, for each task $A_i(t)$, the total workload satisfies $\alpha_i(t)+\beta_{i}(t)+\gamma_{i}(t)=1$.
Consequently, the total workload distribution of AIoT $i$ can be expressed as
\begin{equation}
A_i(t) = \underbrace{\alpha_i(t) A_i(t)}_{\text{AIoT}}  + \underbrace{\beta_{i}(t) A_i(t)}_{\text{UAV}}  + \underbrace{\gamma_{i}(t) A_i(t)}_{\text{GS}}.
\end{equation}

\subsubsection{AIoT based Computing}
$f_i^s$ is defined as the computation resource of AIoT $i$, 
$Q_i^s(t)$ denotes the current queue size of AIoT $i$ at time $t$, and $\alpha_i(t) A_i(t)$ represents the allocated computation task. Therefore, the local computation time of $A_i(t)$ is 
\begin{equation}
T_{i}^{s}(t) = \left(\frac{Q_i^s(t)c_i(t)}{f_i^s}+\frac{\alpha_i(t) d_i(t) c_i(t)}{f_i^s}\right).
\end{equation}

\subsubsection{UAV based Computing}
If the computational resources of the AIoT cannot handle the entire computational task, the AIoT can offload part of the task to the UAV. Due to the limited computational resources of the UAV, it results in higher computational delay but lower transmission delay. Therefore, we define the computational resource of UAV $j$ as $f_j^u$, and the delay of the allocated computation task $\beta_{i}(t) A_i(t)$ on the UAV is represented as

\begin{equation}
\begin{aligned}
T_{i,j}^{u}(t) = p_{i,j}(t) \bigg(\frac{Q_j^u(t)c_i(t)}{f_j^u}&+\frac{\beta_{i}(t)d_i(t)}{R_{i,j}^{a2u}(t)}
\\ &\qquad+\frac{\beta_{i}(t) d_i(t) c_i(t)}{f_j^u}\bigg),
\end{aligned}
\end{equation}
where $Q_j^u(t)$ denotes the queue size of the current UAV $j$ at time $t$.

\subsubsection{GS based Computing}
The GS has powerful computation resources that can reduce computation delay, but it has long transmission delay. Therefore, we define the computation resource of GS $k$ as $f_k^g$, and the delay of the allocated computation task $\gamma_{i}(t) A_i(t)$ on the GS is
\begin{equation}
\begin{aligned}
T_{i,k}^{g}(t) = q_{i,k}(t) \bigg(\frac{Q_k^g(t)c_i(t)}{f_k^g}&+\frac{\gamma_{i}(t)d_i(t)}{R_{i,k}^{a2g}} 
\\ &\qquad+\frac{\gamma_{i}(t) d_i(t) c_i(t)}{f_k^g}\bigg),
\end{aligned}
\end{equation}
where $Q_k^g(t)$ represents the queue size of the current GS $k$ at time $t$.

Besides, at each time slot $t$, the dynamic changes of the task queues for each AIoT, UAV, and GS, denoted as $Q_i^s(t)$, $Q_j^u(t)$, and $Q_k^g(t)$, i.e., 
\begin{equation}
Q_i^s(t+1) = \max \{0,   Q_i^s(t) + \alpha_i(t) d_i(t) - \tau f_i^s\},
\end{equation} 
\begin{equation}
Q_j^u(t+1) = \max\{0,   Q_j^u(t) + \sum_{i=1}^{I} p_{i,j}(t) \beta_{i}(t)d_i(t) - \tau f_j^u\},
\end{equation}
and
\begin{equation}
Q_k^g(t+1) = \max\{0,  Q_k^g(t) + \sum_{i=1}^{I} q_{i,k}(t) \gamma_{i}(t)d_i(t) - \tau f_k^g\}.
\end{equation}

As above, tasks can be processed on AIoTs, UAVs and GSs.	
Hence, the total delay for AIoT $i$ tasks in time slot $t$ is calculated as
\begin{equation}
T_{i}^{t}(t) = T_{i}^{s}(t) + \sum_{j=1}^J T_{i,j}^{u}(t) + \sum_{k=1}^K T_{i,k}^{g}(t).
\end{equation}

Consequently, the cumulative time cost for all AIoT-generated tasks in $T$ is 
\begin{equation}
\Phi(t) = \sum_{t=1}^{T}\sum_{i=1}^I T_{i}^{t}(t).
\end{equation}

\subsection{Problem Formulation}
The optimization problem $\mathscr{P}0$ is formulated to minimize the total execution time by jointly optimizing the flight trajectories of UAVs (i.e., 
$\pmb{L} = \{\theta_j(t), k_j(t)| \forall j \in \mathbb{U}, t \in \mathbb{T}\}$), offloading decisions (i.e., $\pmb{P} = \{p_{i,j}(t)| \forall i \in \mathbb{S}, j \in \mathbb{U}, t \in \mathbb{T}\}, \pmb{Q} = \{q_{i,k}(t)| \forall i \in \mathbb{S}, k \in \mathbb{G}, t \in \mathbb{T}\}$), and the task execution ratios (i.e., $\pmb{\alpha} = \{\alpha_i(t)| \forall i \in \mathbb{S}, t \in \mathbb{T}\}, \pmb{\beta} = \{\beta_i(t)| \forall i \in \mathbb{S}, t \in \mathbb{T}\}, \pmb{\gamma} = \{\gamma_i(t)| \forall i \in \mathbb{S}, t \in \mathbb{T}\}$), i.e., 
\begin{subequations}
\begin{align}
\mathscr{P}0: &\min_{\pmb{L},\pmb{P},\pmb{Q},\pmb{\alpha},\pmb{\beta},\pmb{\gamma}}\sum_{t=1}^{T}\sum_{i=1}^I T_{i}^{t}(t) \notag\\
&\text{s.t. } (\ref{e6}),(\ref{e7}),(\ref{e13})-(\ref{e18}). \notag\\
\end{align}
\end{subequations}

Note that $\mathscr{P}0$ is an MINLP complex problem, as it includes a non-convex objective function and discrete variables. Additionally, the dynamic positions of AIoTs and UAVs, random task arrivals, and uncertain resources available in each time slot increase the dynamic characteristics, which poses significant challenges for traditional optimization algorithms.

\section{Algorithm Design}\label{s4}
In this section, we formalize the problem as a Markov game and propose the GAI-HAPPO algorithm to address the challenges of task offloading and UAV trajectory optimization. 
At the core of our approach is GAI, which combines the strengths of transformer-based actor networks and GAN-based critic networks to enhance the stability and performance of the MARL framework. 
By leveraging GAI, the algorithm overcomes key limitations of traditional RL methods, such as instability and sample inefficiency. The transformer-based actor network generates optimal actions, while the GAN-based critic network refines value estimates through adversarial training. This dual process enables agents to learn optimal policies, while continuously adapting to dynamic environmental conditions. As a result, GAI enhances agent collaboration, stabilizes learning, and improves the efficiency of task offloading and UAV trajectory optimization.

\subsection{Markov Game Formulation} \label{s41}
In the formulated problem $\mathscr{P}0$ in Section \ref{s3}, multiple AIoTs and UAVs coexist, each type possessing its own optimized observation and action spaces.
Therefore, the problem can be modeled as a Markov game.
AIoTs and UAVs can be regarded as $N=I+J$ agents (where $N$ represents the total number of agents), and the problem is represented by the tuple $\langle \mathcal{S}, \{\mathcal{O}_i^s, \mathcal{O}_j^u\}_{i \in \mathbb{S}, j \in \mathbb{U}}, \{\mathcal{A}_i^s, \mathcal{A}_j^u\}_{i \in \mathbb{S}, j \in \mathbb{U}}, \mathcal{R}, \gamma \rangle$.

\subsubsection{State Space}
The state space represents the comprehensive set of variables that define the current status of the system at time $t$, i.e.,
\begin{equation}
\begin{aligned}
s(t) = \{l_i^s(t), &l_j^u(t), l_k^g(t),\\
&Q_i^s(t),Q_j^u(t),Q_k^g(t)|i \in \mathbb{S}, j \in \mathbb{U}, k \in \mathbb{G}\}.
\end{aligned}
\end{equation}

\subsubsection{Observation Space}
Each agent (AIoT or UAV) has its own observation space.

\paragraph{AIoT agent}
The AIoT agent focuses on local computation and task offloading. Its observation space is
\begin{equation}
\begin{aligned}
o_i^s(t) = \{l_i^s(t), l_j^u(t), &l_k^g(t), \\
&Q_i^s(t),Q_j^u(t),Q_k^g(t) | i \in \mathbb{S}, j \in \mathbb{U}, k \in \mathbb{G}\}.
\end{aligned}
\end{equation}

\paragraph{UAV agent}
The UAV agent observation space is
\begin{equation}
\begin{aligned}
o_j^u(t) = \{l_i^s(t), &l_j^u(t), l_k^g(t), \\
&Q_i^s(t),Q_j^u(t),Q_k^g(t) | i \in \mathbb{S}, j \in \mathbb{U}, k \in \mathbb{G}\}.
\end{aligned}
\end{equation}

\subsubsection{Action Space}
The action space defines the set of possible actions an agent can take at time $t$.
\paragraph{AIoT agent}
The action space of the AIoT agent is represented as
\begin{equation}
\begin{aligned}
a_i^s(t) = \{&p_{i,j}(t), q_{i,k}(t), \\
&\alpha_i(t),\beta_i(t), \gamma_i(t) | i \in \mathbb{S}, j \in \mathbb{U}, k \in \mathbb{G} \},
\end{aligned}
\end{equation}
where the AIoT needs to choose which UAV and GS to offload the tasks, as well as how to allocate the task proportions. 

\paragraph{UAV agent}
The action space of the UAV agent is
\begin{equation}
a_j^u(t) = \{\theta_j(t), k_j(t)\},
\end{equation}
where the UAV needs to determine the flying azimuth and distance.

\subsubsection{Reward Function}
The reward function is based on the time required to process a single bit, considering the uncertainty of tasks and the complex variability of the environment. In detail, the reward function $r(t)$ can be expressed as
\begin{equation}
r(t) = \sum_{i=1}^{I}d_i(t)/T_i^t(t).
\end{equation}

The on-policy RL methods use the current policy to gather transitions for updating the value function. 
However, on-policy techniques have several limitations, including issues with sample inefficiency and the need for repeated interactions with the environment to achieve optimal results.
Furthermore, the on-policy RL often experiences policy oscillations and instability, leading to sub-optimal strategies \cite{On_2024}.

\begin{figure}[!t]
\centering
\includegraphics[width=7cm]{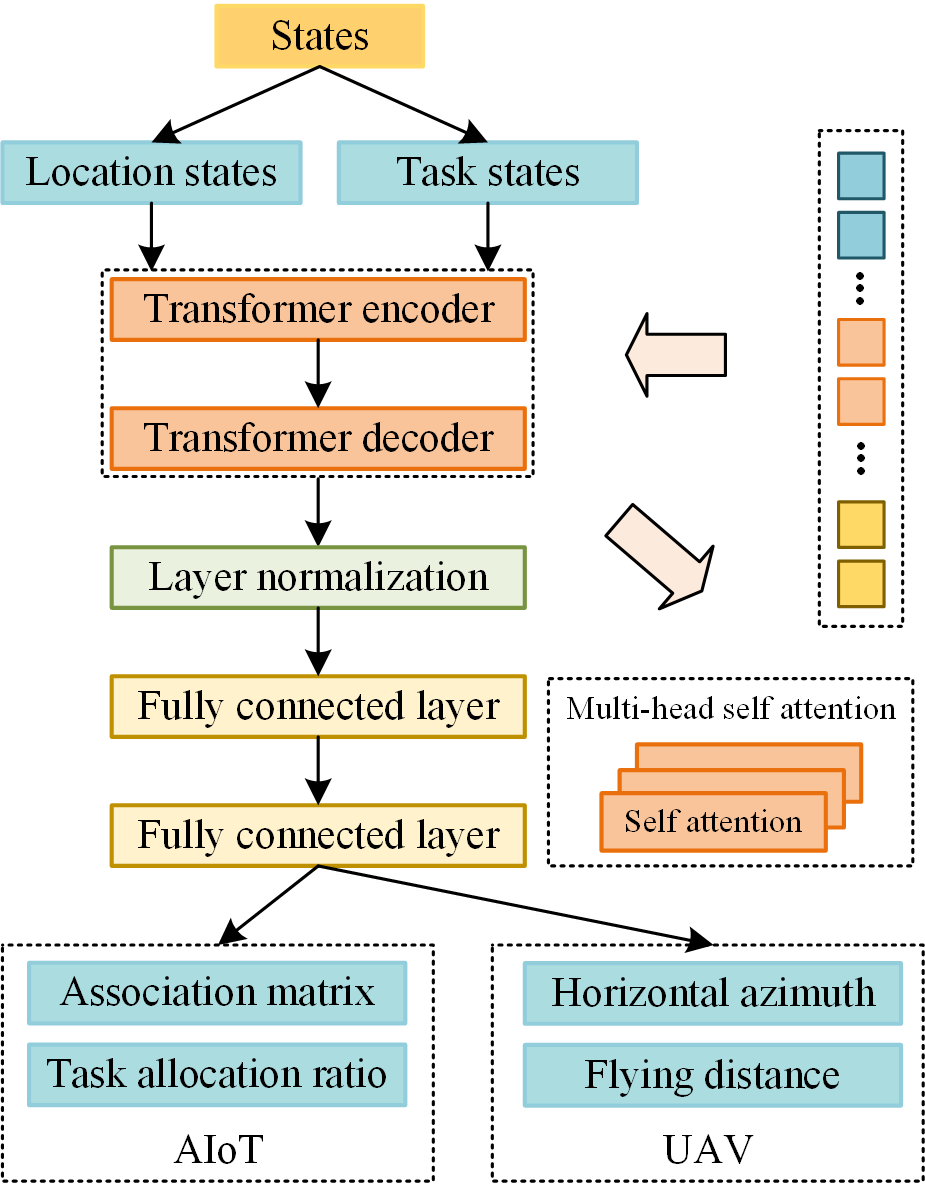}
\caption{Transformer-based actor network.}
\label{f3}
\end{figure}

\subsection{Transformer-based Actor Network}\label{s42}

In the GAI-HAPPO framework, the transformer-based actor network serves as a core generative component that enhances policy learning by capturing structured spatial-temporal dependencies across heterogeneous agents. Leveraging the self-attention mechanism and multi-head architecture, the transformer effectively models long-range interactions in dynamic and partially observable MEC environments. This design enables the actor to focus on salient features across sequential inputs, supporting more robust task offloading and UAV trajectory decisions. Inspired by GAI architectures such as GPT and decision transformer, the transformer module improves the adaptability and generalization of the learned policies under uncertainty.

The core mechanism of the actor network is the self-attention mechanism in transformers, which enables the agent to focus on relevant features from complex input data, thereby making informed decisions. 
Consequently, the input data $Z$, are used to generate the query matrix $Q$, key matrix $K$, and value matrix $V$ through linear transformations, i.e., 
\begin{equation}
Q=Z W^Q, \quad K=Z W^K, \quad V=Z W^V,
\end{equation}
where $W^Q$, $W^K$, and $W^V$ are the weight matrices for the query, key, and value, respectively.
Subsequently, the self-attention mechanism, crucial for recognizing relevant information across various positions in the data, is defined as:	
\begin{equation}
\operatorname{Att}(Q, K, V)=\operatorname{softmax}\left(\frac{Q K^T}{\sqrt{d_k}}\right) V,
\end{equation}
where $Q$, $K$, and $V$ are the query, key, and value matrices computed from the input data, respectively.
The dimensionality of the key vectors is denoted by $d_k$, which influences the scaling factor in the attention mechanism.

Furthermore, the actor network employs multi-head attentions, which allows the model to focus on different aspects of the input data simultaneously. 
The multi-head attention divides the $K$, $Q$, and $V$ vectors into several sub-vectors of equal length, forming sub-spaces that allow the model to concentrate on various aspects of information, thus generating head vectors as
\begin{equation}
\text {head}_i=\operatorname{Att}\left(Q W_i^Q, K W_i^K, V W_i^V\right).
\end{equation}
The outputs from all heads are concatenated and a linear transformation $W^O$ is applied. Therefore, the multi-head attention process is defined as
\begin{equation}
\operatorname{MH}(Q, K, V)=\operatorname{Con}\left(\text{head}_1, \text{head}_2, \ldots, \text {head}_h\right) W^O.
\end{equation}

As depicted in Fig. \ref{f3}, the input to the transformer-based actor network comprises two main components: location states and task states. 
These inputs are initially processed by the transformer encoder and decoder, responsible for extracting relevant features and capturing relationships between different states. The multi-head self-attention mechanism enables the model to attend to various parts of the data simultaneously, enhancing its capability to comprehend complex interactions within the environment. After processing by the encoder and decoder, the data undergoes layer normalization to stabilize and accelerate the training process. The normalized output is then fed into fully connected layers, which further process the data to generate meaningful action probabilities.
The output from the fully connected layers branches into two main pathways. The network generates an association matrix and task allocation ratio, determining how tasks are distributed among the AIoTs and their respective allocation proportions. The network also outputs horizontal azimuth and flying distance, crucial for optimizing UAV trajectories.

By incorporating the multi-head attention and layer normalization, the transformer-based actor network is able to handle the dynamic and heterogeneous nature of the multi-agent environment more effectively. The network can attend to multiple aspects of the state space simultaneously and capture long-term dependencies between task offloading and UAV trajectory optimization. 
This results in highly accurate and stable decision-making policies, empowering UAVs and AIoTs to optimize their trajectories and task assignments effectively within the GAI-HAPPO framework.

\subsection{GAN-based Critic Network}\label{s43}

In GAI-HAPPO, the critic network adopts a generative adversarial architecture that improves value estimation under uncertainty. Traditional critics in MARL often suffer from instability and high variance due to the noise from agent interactions and the non-stationary nature of decentralized environments. To address this issue, we introduce a GAN-based critic in which the generator produces value estimates and the discriminator evaluates their plausibility by comparing them with observed returns. This adversarial mechanism establishes a dynamic feedback process where the generator continuously learns to produce more accurate return distributions while the discriminator discourages unrealistic or poorly generalized predictions. This design enhances the robustness of value learning and acts as an implicit regularizer, mitigating issues such as gradient vanishing and overfitting.

In this setup, the generator network estimates state values, while the discriminator network evaluates the quality of these estimated values. The adversarial nature of  encourages the generator to refine its value estimates continually, making them more reliable. This dual process helps stabilize the critic network, mitigating issues typically found in on-policy RL methods and enhancing the overall performance of the GAI-HAPPO algorithm.

\begin{figure}[!t]
\centering
\includegraphics[width=8cm]{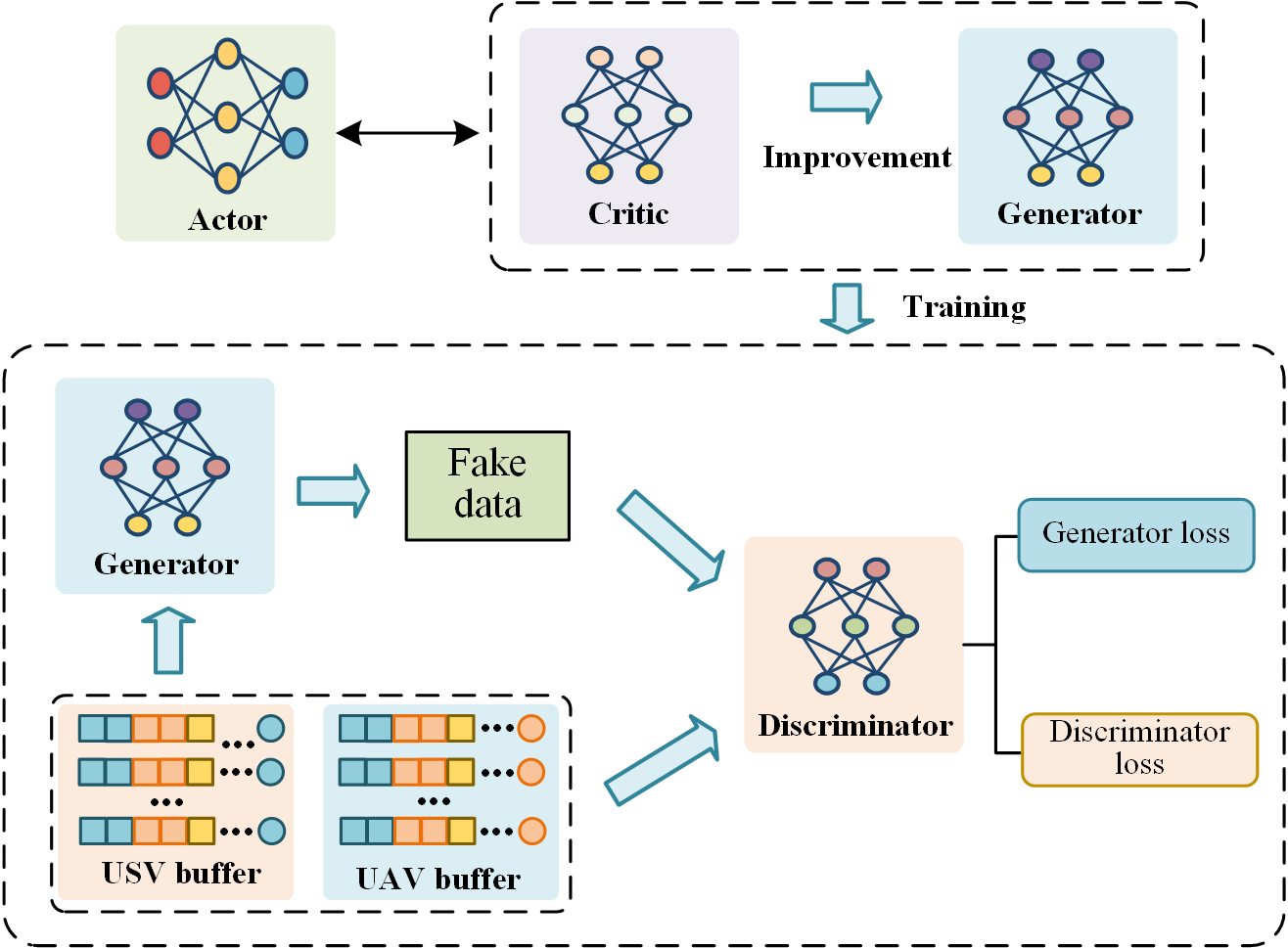}
\caption{GAN-based critic network.}
\label{f4}
\end{figure}

As shown in Fig. \ref{f4}, by integrating GAN, the critic network benefits from the adversarial training mechanism, where the generator and discriminator work in tandem to refine the action-value estimates.
The generator network $G$ produces estimated state values $\hat{V}(s_t)$:
\begin{equation}
\hat{V}\left(s_t\right)=G\left(s_t , \theta_G\right).
\end{equation}  

The discriminator network $D$ outputs a probability $D\left(s_t, V\right)$ that a given value is a true state value. Then, the loss for the discriminator network is calculated as
\begin{equation}
\begin{split}
L_D=-\mathbb{E}\big[\log D(s_t,& V^*(s_t))\big]\\
&-\mathbb{E}\big[\log (1-D(s_t, \hat{V}(s_t)))\big].
\end{split}
\end{equation} 
The loss function aims to maximize the likelihood that the discriminator correctly identifies true state values, while minimizing the likelihood of incorrectly classifying generated state values as true.

The loss for the generator network is calculated as
\begin{equation}
L_G=\mathbb{E}\left[\log \left(1-D\left(s_t, \hat{V}\left(s_t\right)\right)\right)\right].
\end{equation}
The loss function encourages the generator to produce state values that the discriminator recognizes as true.

By integrating the GAN-based critic network, the GAI-HAPPO framework can achieve more stable and accurate value estimations, which is crucial for optimizing the collaboration in task offloading and trajectory.

\begin{figure}[!t]
\centering
\includegraphics[width=8cm]{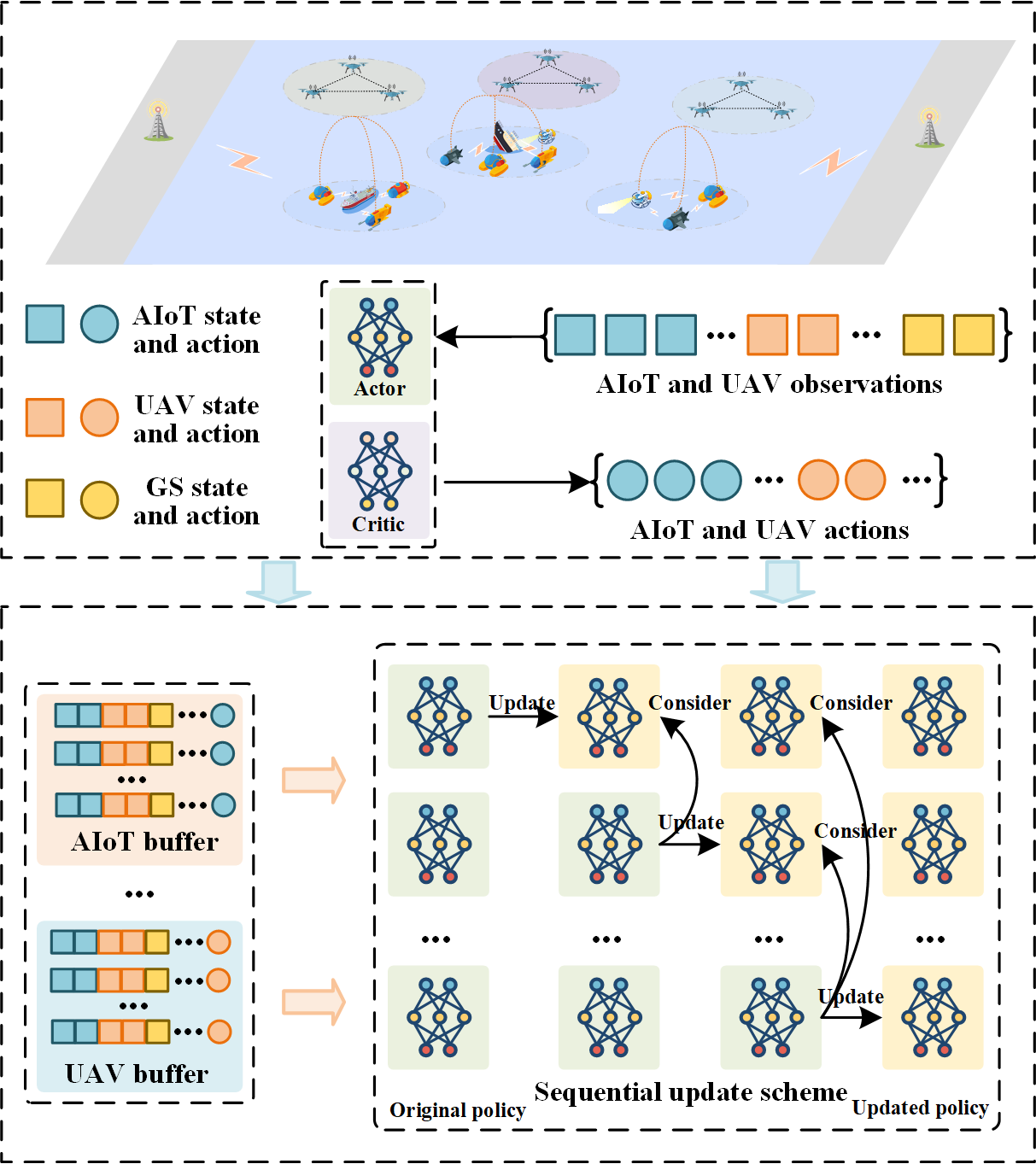}
\caption{The GAI-HAPPO framework in cooperative UAV-GS MEC.}
\label{f2}
\end{figure}

\subsection{GAI-HAPPO}\label{s44}
Algorithm \ref{algorithmic1} leverages GAI to enhance MARL via a coordinated policy optimization process. 
As illustrated in Fig. \ref{f2}, GAI-HAPPO uses a sequential update scheme to improve the collaboration in heterogeneous multi-agent systems \cite{HARL_2024}.	
Each agent policy $\bm{\pi}^i_{\theta^i}$ is parameterized by $\theta^i$, creating a joint policy $\bm{\pi}$ parameterized by $\bm{\theta} = \{\theta^1,...,\theta^n\}$.
During each iteration $k + 1$, with a permutation of agents $i_{1: n}$, each agent $i_{m}$ (where $m \in \{1,...,n\}$) sequentially optimizes its policy parameter.	
This optimization involves agent $i_m$ selecting a policy parameter $\theta_{k+1}^{i_m}$ that maximizes the following objective, i.e.,
\begin{equation}
\begin{split}
&\mathcal{L}(\theta_{k+1}^{i_m}) = \mathbb{E}_{s \sim \rho_{\bm{\pi}_{\bm{\theta}^k}}, \bm{a}\sim \bm{\pi}_{\bm{\theta}_{k}}}   \bigg[ \min \bigg( \frac{\pi_{i_m}^{\theta_{i_m}}(a_{i_m}|s)}{\pi_{i_m}^{\theta_{i_m}^k}(a_{i_m}|s)} M^{i_{i:m}}(s, \bm{a}), \\
&\qquad\qquad\qquad \text{clip}\bigg( \frac{\pi_{i_m}^{\theta_{i_m}}(a_{i_m}|s)}{\pi_{i_m}^{\theta_{i_m}^k}(a_{i_m}|s)}, 1 \pm \epsilon \bigg)M^{i_{i:m}}(s, \bm{a})\bigg)\bigg],
\end{split}\label{eq33}
\end{equation}
where
\begin{equation}
M^{i_{1: m}}(s, \bm{a})=\frac{\boldsymbol{\pi}_{\boldsymbol{\theta}_{k+1}^{i_{1: m-1}}}^{i_{1: m-1}}\left(\bm{a}^{i_{1: m-1}} \mid s\right)}{\boldsymbol{\pi}_{\boldsymbol{\theta}_k^{i_{1: m}}}^{i_{1:m-1}}\left(\bm{a}^{i_{1: m-1}} \mid s\right)} \hat{A}(s, \bm{a}).
\label{eq34}
\end{equation}
$\hat{A}(s, \bm{a})$ is the advantage function in PPO. 
$M^{i_{1 m}}$ represents the advantage function in GAI-HAPPO, which needs to consider the parameters of the other agents. 
$\epsilon$ denotes a hyperparameter used to control the magnitude of the policy updates.
Notably, (\ref{eq34}) aligns well with the sequential update scheme in HAPPO. For agent $i_m$, since previous agents $i_{1:m-1}$ have already updated, the compound policy ratio for $M^{i_{1:m}}$ is easy to obtain.

The generator network is updated to minimize the following loss function $\theta_G$, effectively replacing the V-value network update, i.e.,
\begin{equation}
\theta_G = \arg \min \frac{1}{B T} \sum_{b=1}^B \sum_{t=0}^T\left(G\left(s_t\right)-\hat{R}_t\right)^2.
\end{equation}

During the policy optimization process, the generator and discriminator networks within the GAN-based critic network are updated to refine the action-value estimates.
The update rules for the generator and discriminator networks are 
\begin{equation}
\theta_G \leftarrow \theta_G-\alpha_G \nabla_{\theta_G} L_G,
\end{equation}
and
\begin{equation}
\theta_D \leftarrow \theta_D-\alpha_D \nabla_{\theta_D} L_D,
\end{equation}
respectively.

Initially, the actor networks for each agent, along with the generator and discriminator networks, are initialized (line \ref{a1}). 
For each episode, agents collect trajectory data by executing the joint policy, and store the observed transitions in a batch (lines \ref{a3}-\ref{a4}). 
Using generalized advantage estimation, the advantage function is calculated to evaluate the quality of actions (line \ref{a5}). 
A random permutation of agents is generated to diversify the update sequence, reducing update bias (lines \ref{a6}-\ref{a7}). 
Each agent policy (actor network) is updated iteratively by optimizing the PPO-Clip objective, using the advantage function as a baseline (lines \ref{a9}-\ref{a10}). 
This process is repeated for all agents, allowing the algorithm to converge toward a stable policy. 
Finally, the generator and discriminator networks are updated to further refine data quality, promoting stability in the MARL framework (line \ref{a12}).
Through this iterative process across multiple episodes, GAI-HAPPO enhances the adaptability and robustness of agents in complex, multi-agent environments.

\begin{algorithm}[!t]
\caption{GAI-HAPPO}
\label{algorithmic1}
\begin{algorithmic}[1]
\REQUIRE Batch size $B$, number of agents $n$, episodes $K$, and steps per episode $T$.
\STATE \textbf{Initialize:} Actor networks $\theta^{i_m}$, generator network $G$, and discriminator network $D$. \label{a1}
\FOR{$k = 0$ to $K-1$}
    \STATE Collect trajectories by running the joint policy $\boldsymbol{\pi}_{\boldsymbol{\theta}_k}=(\pi_{\theta_k^1}^1, \ldots, \pi_{\theta_k^n}^n)$. \label{a3}
    \STATE Store transitions $\{(s_t, o_t^i, a_t^i, r_t, s_{t+1}, o_{t+1}^i),$ $\forall i \in N, t \in T\}$ in $B$. \label{a4}
    \STATE Calculate the advantage function $\hat{A}(\mathrm{s}, \mathbf{a})$ using generalized advantage estimation.\label{a5}
    \STATE Generate a random permutation of agents $i_{1: n}$.\label{a6}
    \STATE Set $M^{i_1}(\mathrm{s}, \mathbf{a})=\hat{A}(\mathrm{s}, \mathbf{a})$.\label{a7}
    \FOR{agent $i_m = i_1, \ldots, i_n$}
        \STATE Update actor $i_m$ with $\theta_{k+1}^{i_m}$, the argmax of the PPO-Clip objective by using (\ref{eq33}).\label{a9}
        \STATE Calculate $M^{i_{1: m+1}}(s, \bm{a})$. \label{a10}
    \ENDFOR
    \STATE Update generator network $G$ and discriminator network $D$. \label{a12}
\ENDFOR
\end{algorithmic}
\end{algorithm}

\section{Simulation Results}\label{s5}
To evaluate the performance of the proposed GAI-HAPPO algorithm, we conduct numerical experiments and assess its effectiveness and superiority against four benchmark algorithms, described as follows.

\makeatletter
\let\TMCsavedsubsubsectionlevel\toclevel@subsubsection
\let\toclevel@subsubsection\toclevel@subsection
\makeatother
\subsubsection{Heterogeneous-Agent Advantage Actor-Critic (HAA2C)}
HAA2C \cite{HARL_2024} expands on the advantage actor critic framework, targeting at the decision-making challenges faced by heterogeneous agents in MARL environments.	
\subsubsection{HAPPO}
HAPPO algorithm \cite{HARL_2024} is an evolution of the foundational PPO algorithm, specifically tailored to handle collaborative behaviors among heterogeneous agents within MARL settings.	
\subsubsection{GAN-HAPPO}
GAN-HAPPO integrates the principles of GAN with the HAPPO framework. 
The combination leverages the strengths of GAN to enhance the stability and accuracy of the critic network within the HAPPO algorithm.
\subsubsection{Transformer-HAPPO}
Transformer-HAPPO incorporates transformer architecture into the HAPPO framework to address the sequential and spatial-temporal complexities in MARL environments. 

\subsubsection{Heuristic Delay-Optimized Offloading (HDOO)}
HDOO is a traditional baseline that allocates each task to the node (UAV or GS) with the lowest estimated execution delay based on current system conditions.

\makeatletter
\let\toclevel@subsubsection\TMCsavedsubsubsectionlevel
\makeatother

\subsection{Parameter Setting}
We consider a scenario within a $1,000m \times 1,000m$ area, where 6 AIoTs are deployed to generate tasks, and 4 UAVs along with 2 GSs are utilized for task processing \cite{Energy_tvt_2024,UAV_2024}.
At each time slot, every AIoT generates a computation-intensive task. 
Subsequently, UAVs move to optimal positions while AIoTs, UAVs, and GSs execute tasks based on predetermined actions.
Detailed parameters are listed in Table \ref{T2} \cite{wang_hybrid_2021, Soft_2023}. 
During simulations, we discretize the time into time steps to ensure precise control and evaluation.

\begin{table}[!t]
\caption{Simulation Parameters}
\centering
\begin{tabular}{|m{1.6cm}<{\centering}|m{1.8cm}<{\centering}||m{1.6cm}<{\centering}|m{2.0cm}<{\centering}|}
\hline
\textbf{Parameter} & \textbf{Value} & \textbf{Parameter} & \textbf{Value} \\ \hline
$\zeta_L$ & $2.3$ & $\zeta_{NL}$ & $34$ \\ \hline
$a$ & $5.0188$ & $b$ & $0.3511$ \\ \hline
$h_i^s$, $h_j^u$ & 0 m, 30 m & $k^{max}$ & 30 m \\ \hline
$C$ & $3\times10^8$ m/s  & $c_i(t)$ & 270 cycles/bit \\ \hline
$P_i^s$ & 1 W & $N_G$ & -114 dBm \\ \hline
$\lambda_i$ & 15 Mbits& Learning rate & $5\times10^{-5}, 10^{-4}$ \\\hline
\end{tabular}
\label{T2}
\end{table}

\begin{figure}[htbp]
\centering
\subfloat[Reward with different activation functions.]{
\includegraphics[width=7cm]{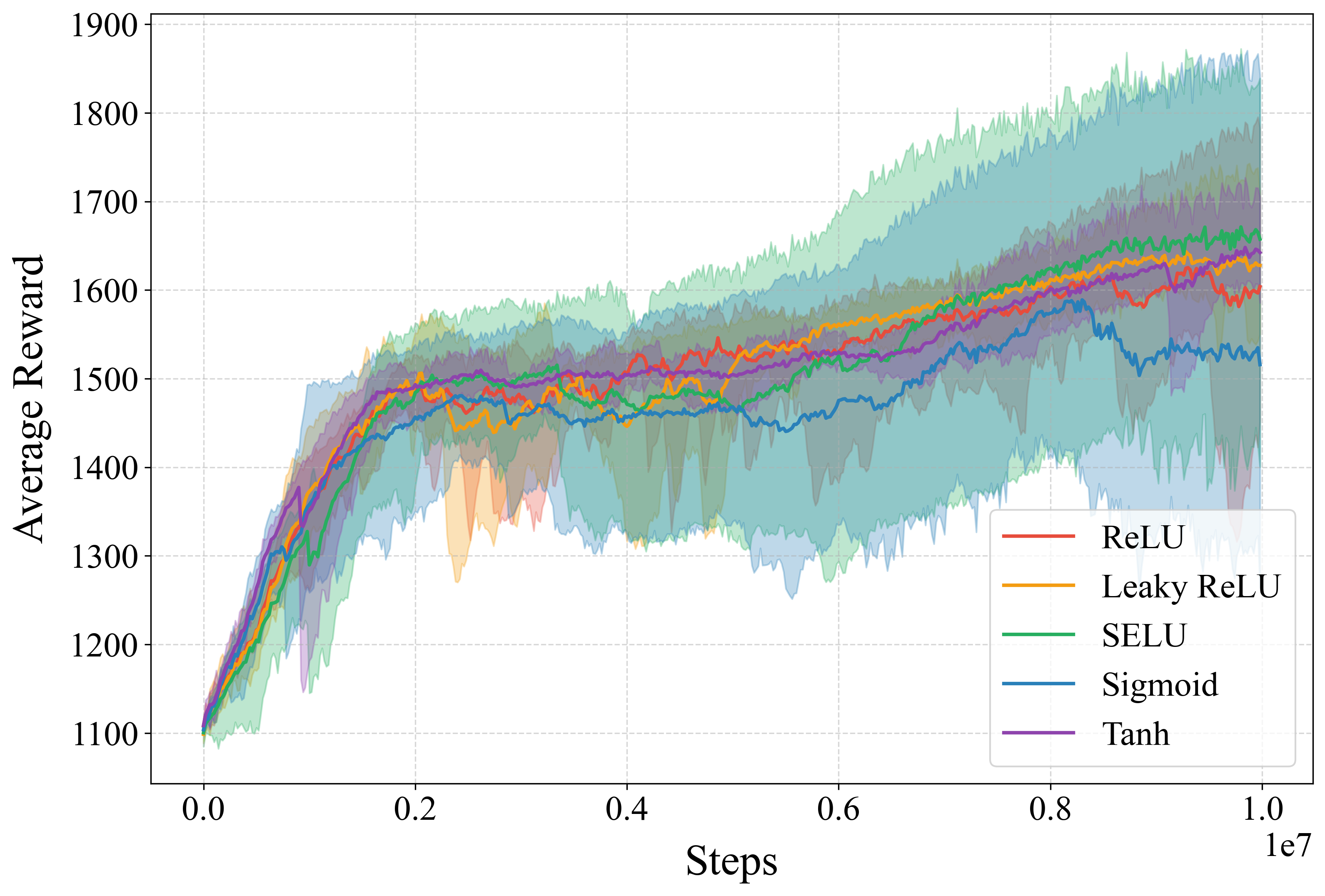}\label{f5a}
}\\
\subfloat[Reward with different actor learning rates.]{
\includegraphics[width=7cm]{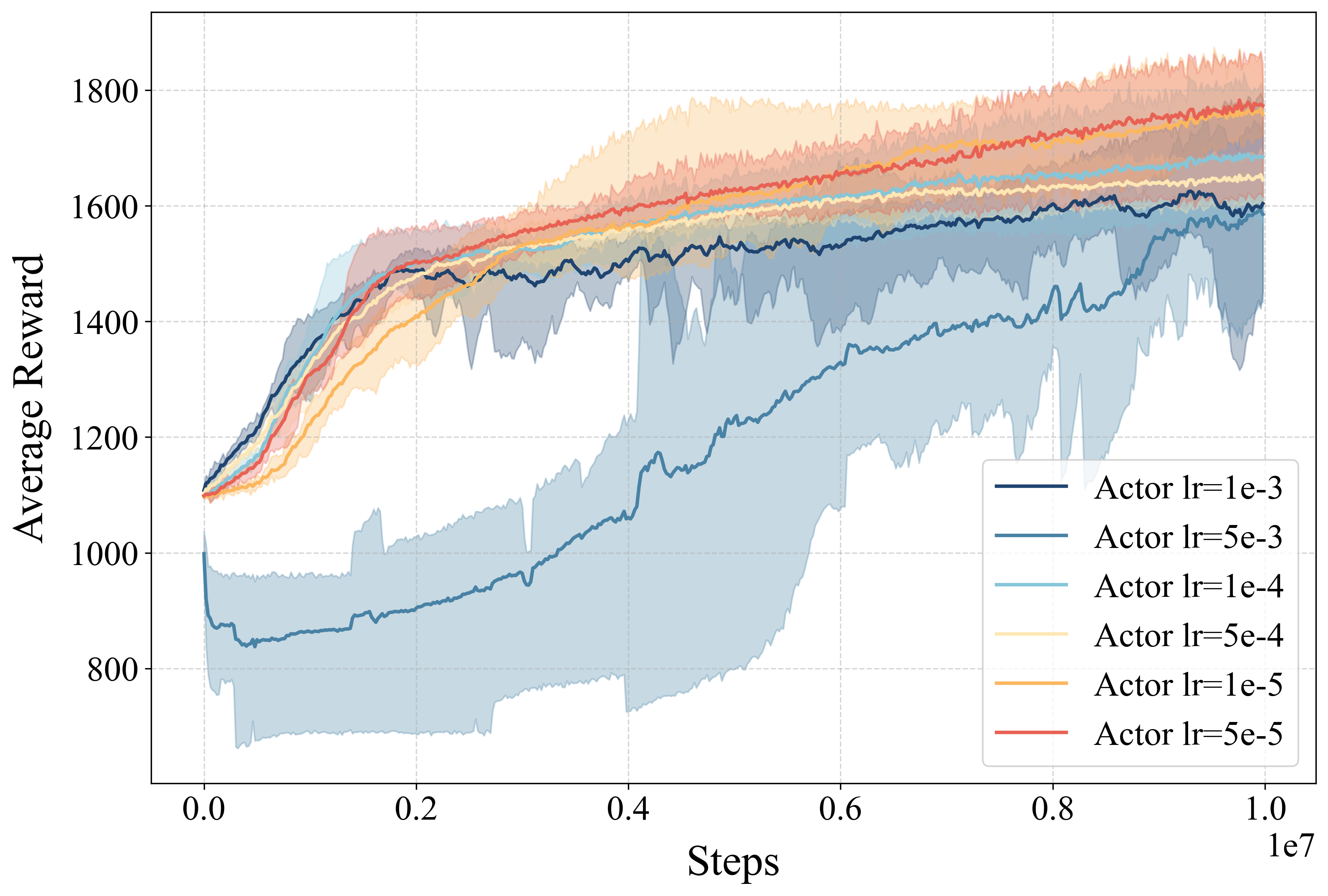}\label{f5b}
}\\
\subfloat[Reward with different critic learning rates.]{
\includegraphics[width=7cm]{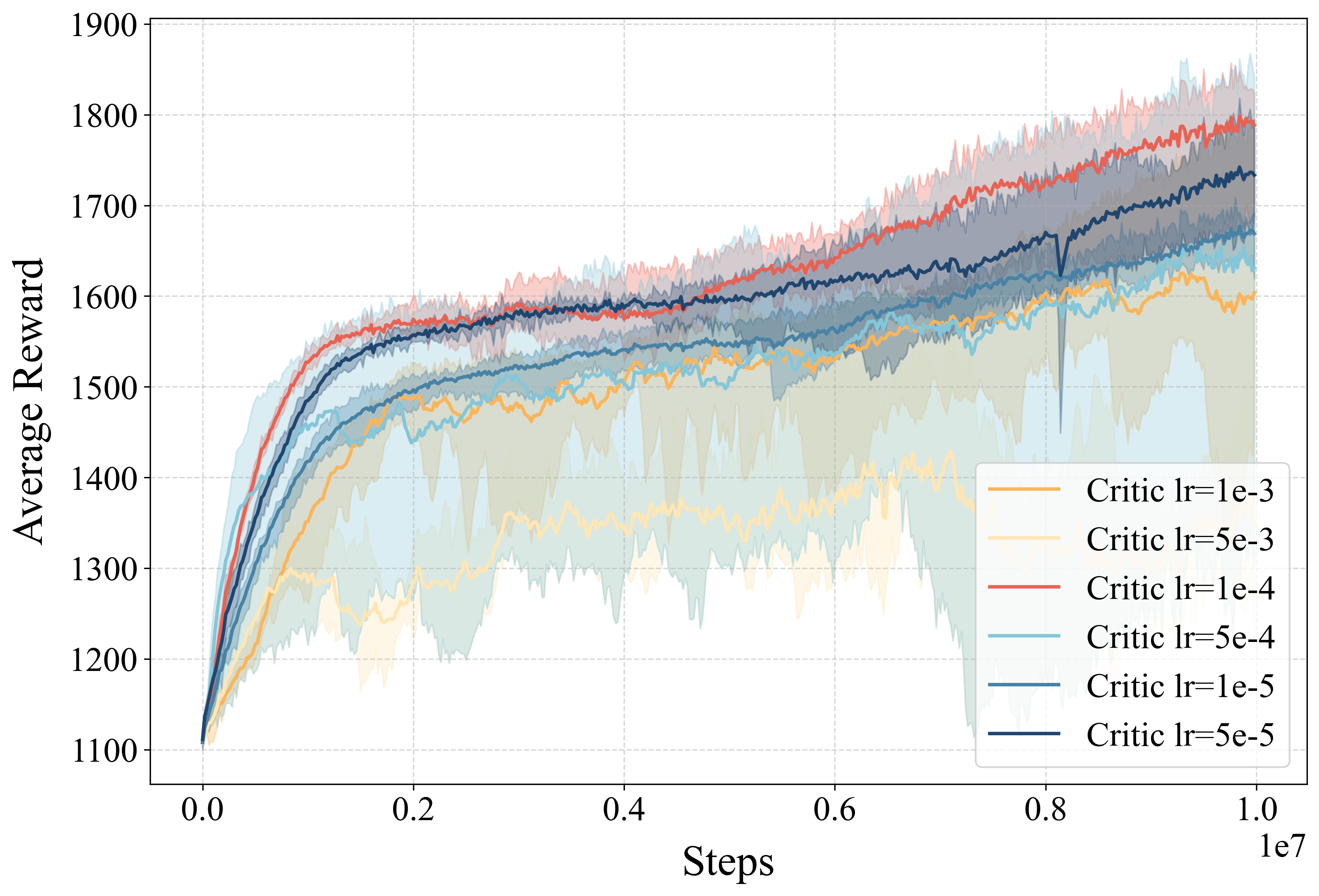}\label{f5c}
}
\caption{Performance with different parameters.}
\label{f5}
\end{figure}

\begin{figure}[!t]
\centering
\includegraphics[width=7cm]{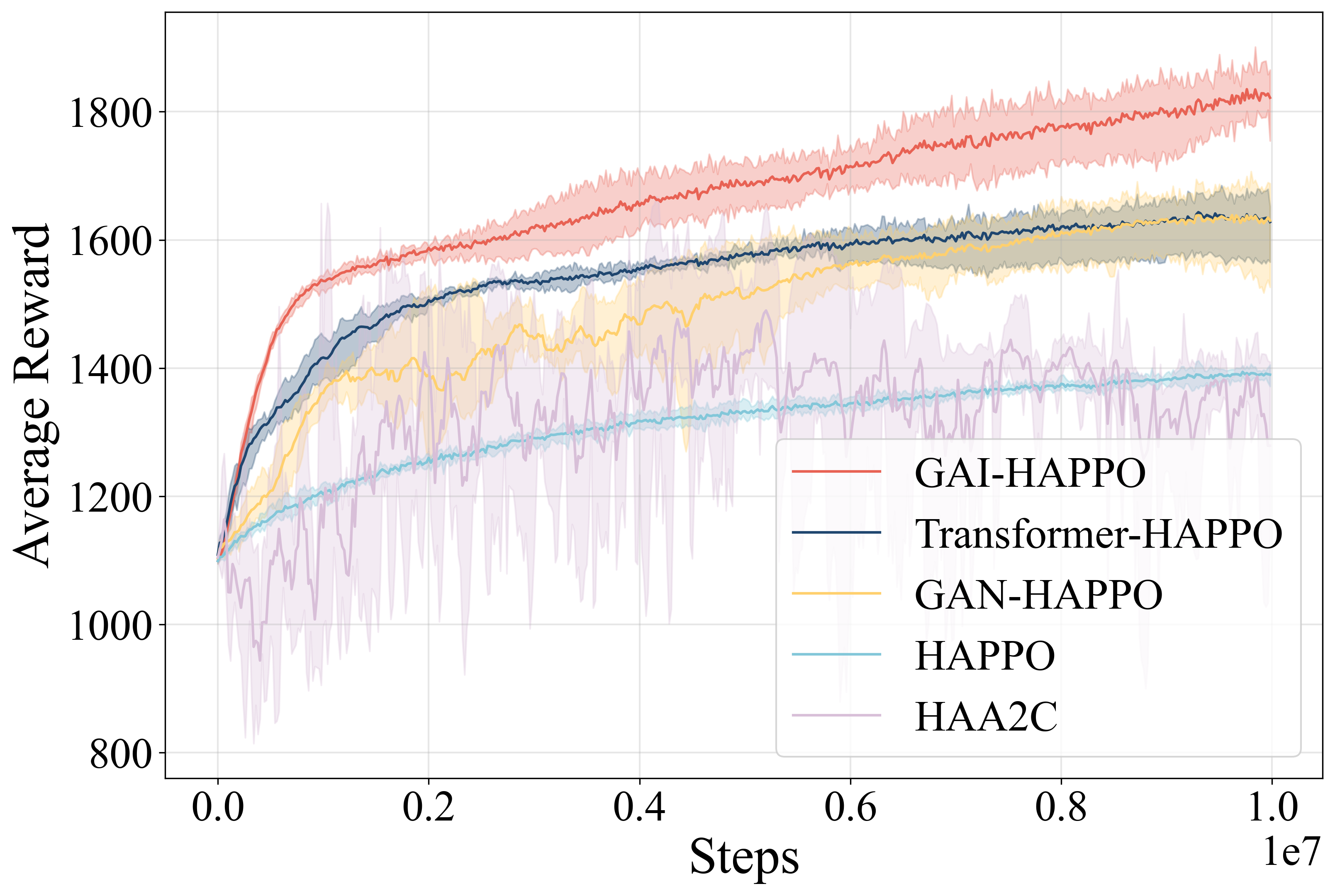}
\caption{Reward with different algorithms.}
\label{f6}
\end{figure}

\begin{figure*}[htbp]
\centering
\subfloat[Average reward.\label{f7a}]{
    \includegraphics[width=5.8cm]{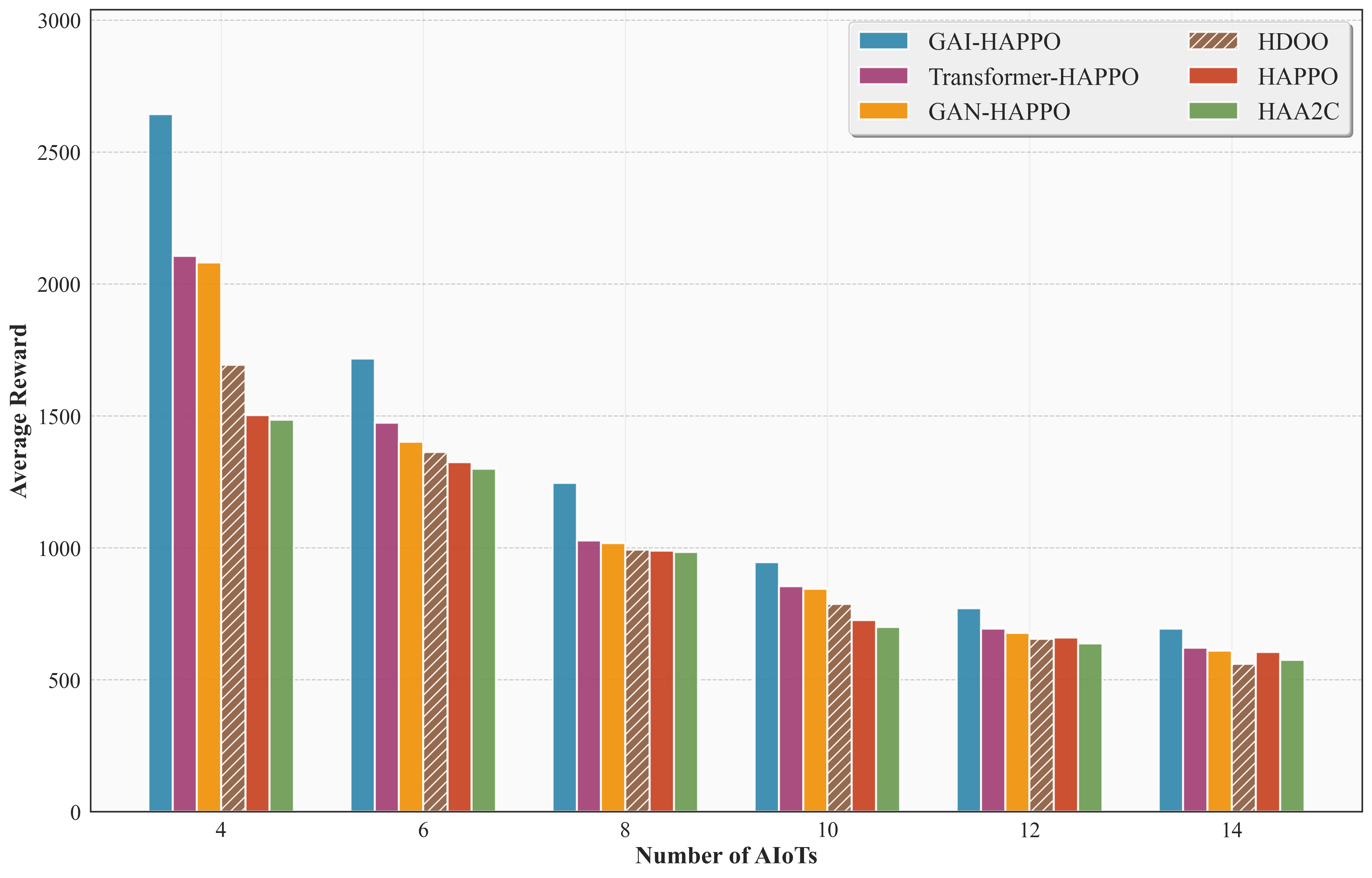}
}
\subfloat[Average execution time.\label{f7b}]{
    \includegraphics[width=5.8cm]{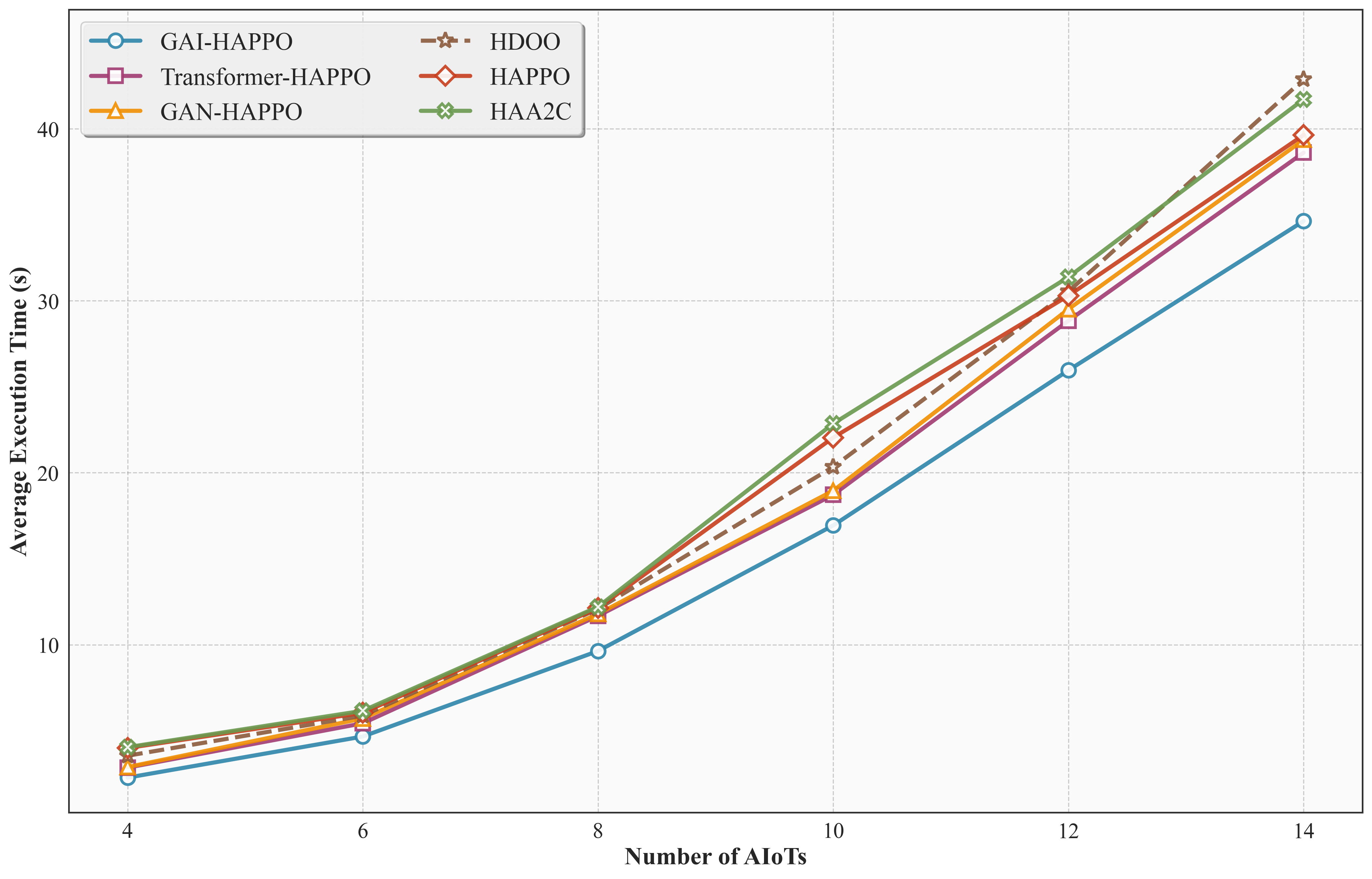}
}
\subfloat[Average computation of ratio.\label{f7c}]{
    \includegraphics[width=5.8cm]{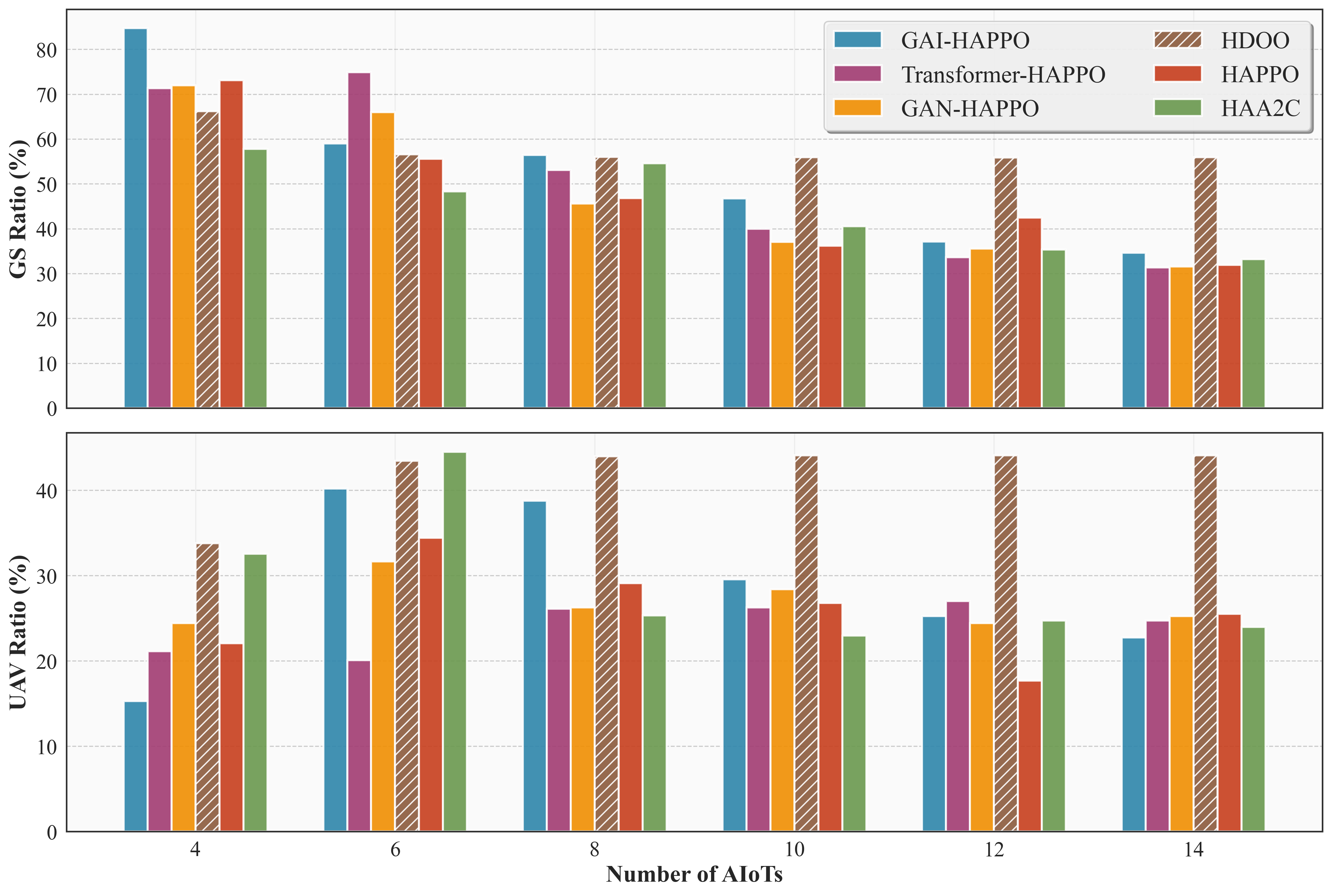}
}
\caption{Performance with different numbers of AIoTs.}
\label{f7}
\end{figure*}

\subsection{Performance with Different Parameters}
We firstly conduct the performance of the average reward with different activation functions, actor learning rates, and critic learning rates, respectively. 
In Fig. \ref{f5}(a), different activation functions show distinct impacts on reward performance and training stability. SELU achieves the highest final reward due to its self-normalizing property, which enhances the network's ability to learn complex policies. However, it also introduces higher variability, indicating potential sensitivity to environmental fluctuations. ReLU yields a smoother convergence with fewer oscillations but results in a lower final reward, due to the limited expressiveness in capturing fine-grained feature dependencies. Sigmoid performs worst in both stability and reward, as its output saturation hinders effective gradient propagation. Leaky ReLU balances convergence stability and final performance, achieving rewards comparable to SELU with improved robustness during training.

In Fig. \ref{f5}(b), different actor learning rates exhibit varying effects on training stability and final performance. Learning rates of $10^{-5}$ and $5\times10^{-5}$ achieve the highest average rewards and maintain stable convergence throughout the training, reflecting effective gradient updates and smooth policy refinement. Slightly larger rates, such as $10^{-4}$ and $5\times10^{-4}$, lead to faster convergence but exhibit moderate fluctuations in early training stages, suggesting increased sensitivity to learning dynamics. In contrast, a high learning rate of $5\times10^{-3}$ results in the poorest performance, with erratic fluctuations and low final reward, due to overly aggressive updates.

In Fig. \ref{f5}(c), different critic learning rates show notable effects on performance. A rate of $10^{-4}$ achieves the best balance, yielding stable convergence and high final rewards. The rate of $10^{-3}$ also facilitates fast convergence, but introduces slight fluctuations, indicating reduced stability. Conversely, smaller rates such as $10^{-5}$ and $5\times10^{-5}$ lead to slow reward growth and higher variability, suggesting insufficient update magnitude for effective learning. These results highlight that moderate learning rates provide reliable training dynamics for the critic network, whereas overly small or large rates may impede convergence efficiency or compromise robustness.

\begin{figure*}[htbp]
\centering
\subfloat[Average reward.\label{f8a}]{
    \includegraphics[width=5.8cm]{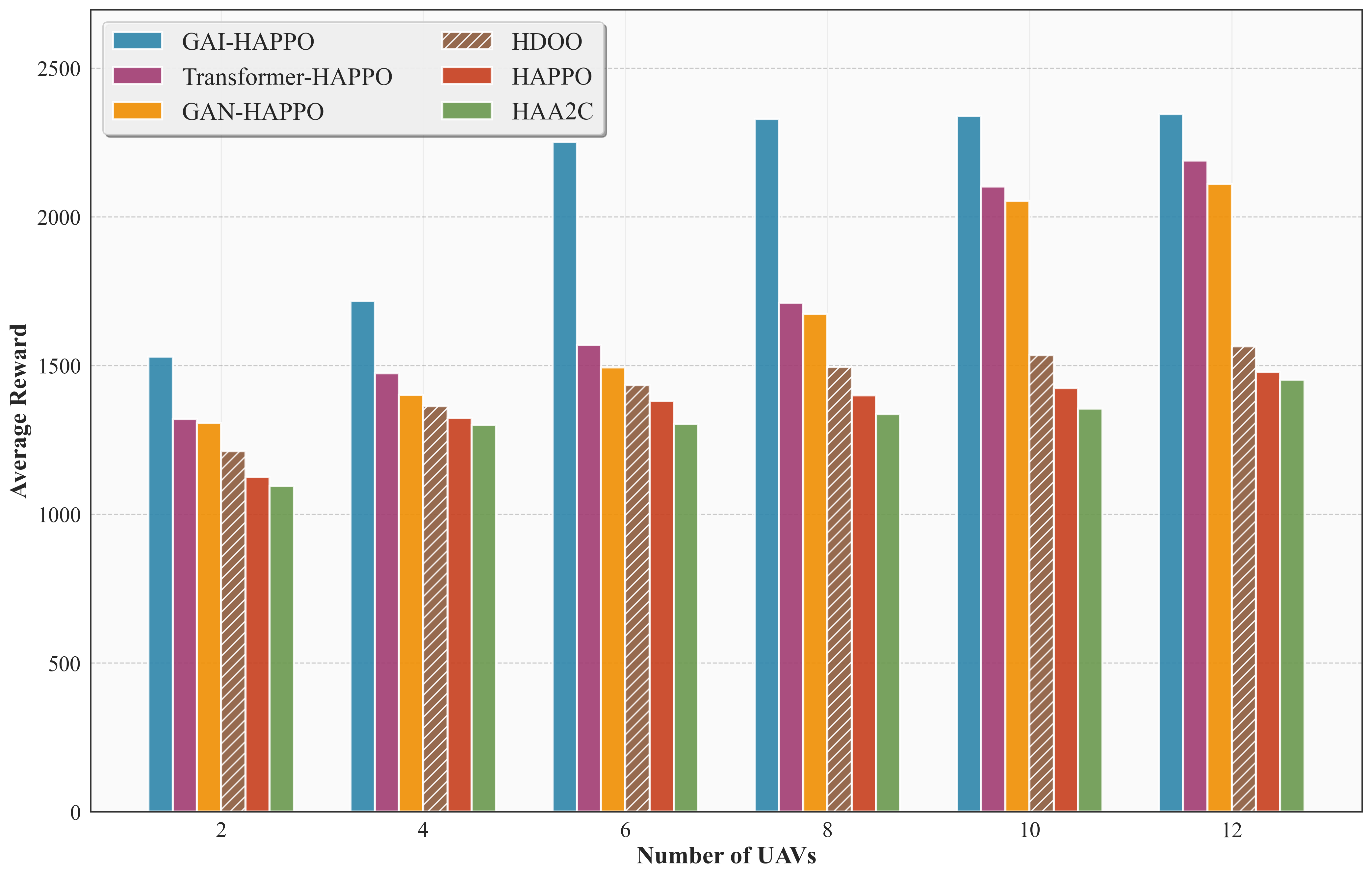}
}
\subfloat[Average execution time.\label{f8b}]{
    \includegraphics[width=5.8cm]{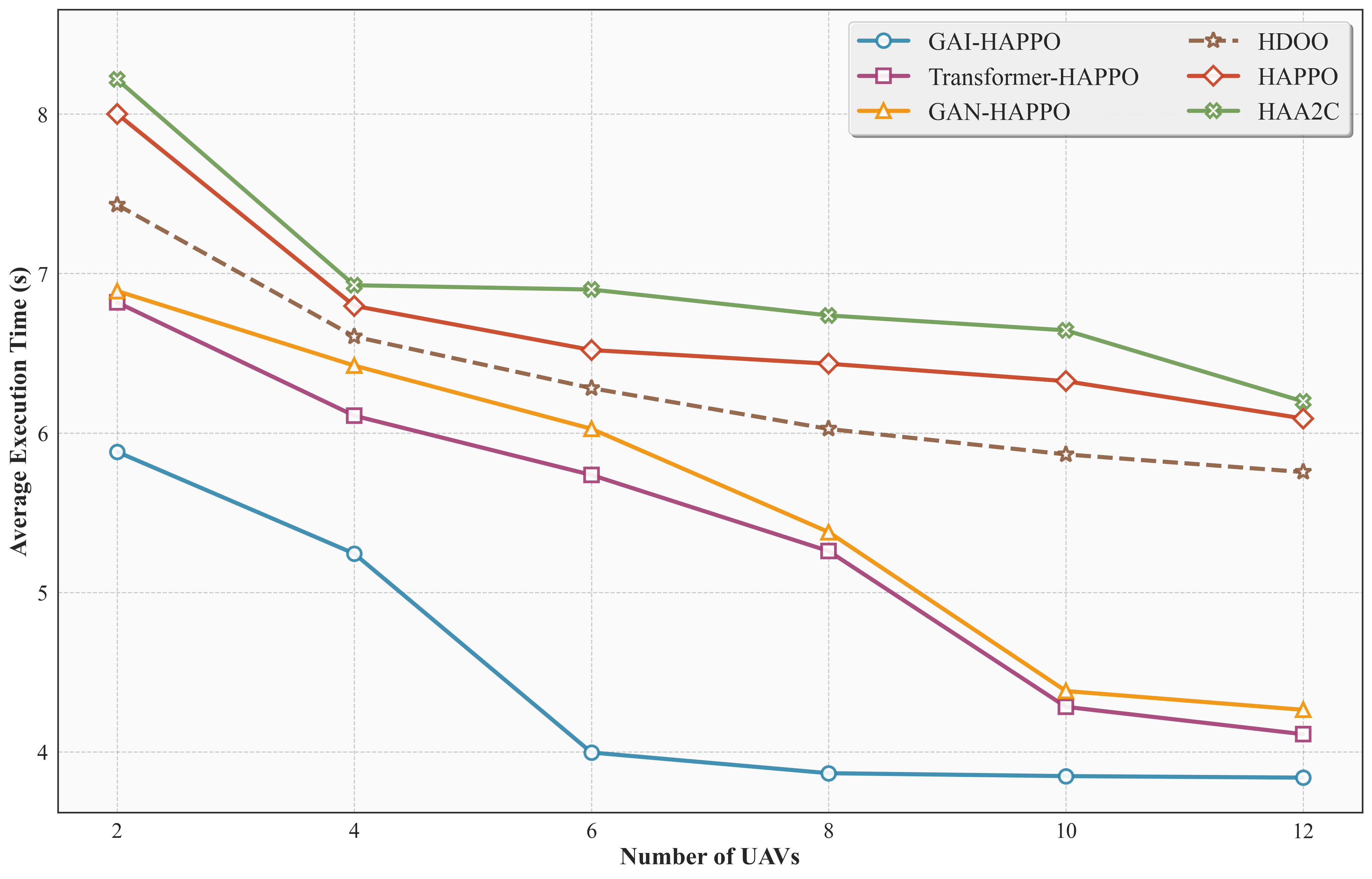}
}
\subfloat[Average computation of ratio.\label{f8c}]{
    \includegraphics[width=5.8cm]{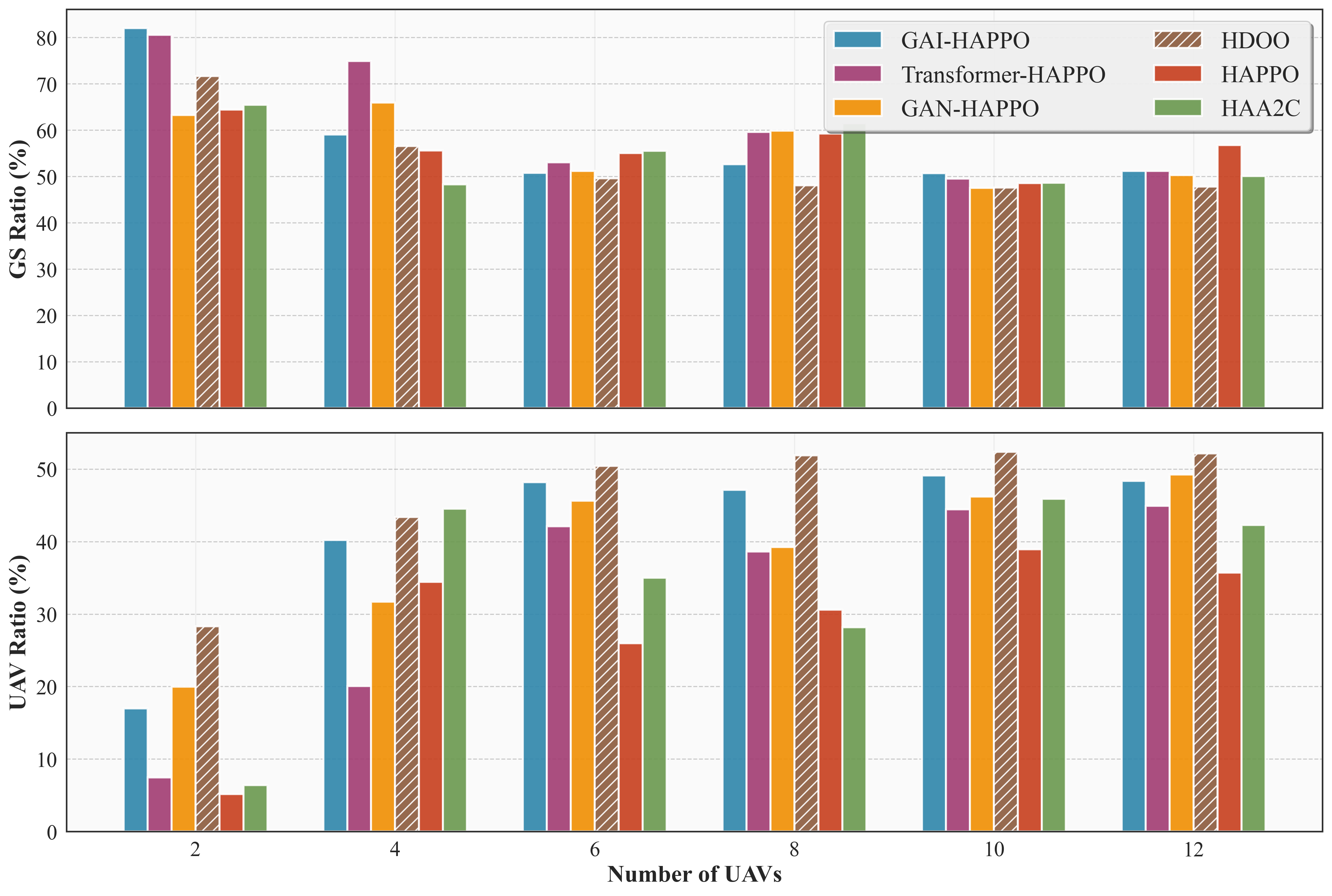}
}
\caption{Performance with different numbers of UAVs.}
\label{f8}
\end{figure*}

\begin{figure*}[htbp]
\centering
\subfloat[Average execution reward.\label{f9a}]{
    \includegraphics[width=5.8cm]{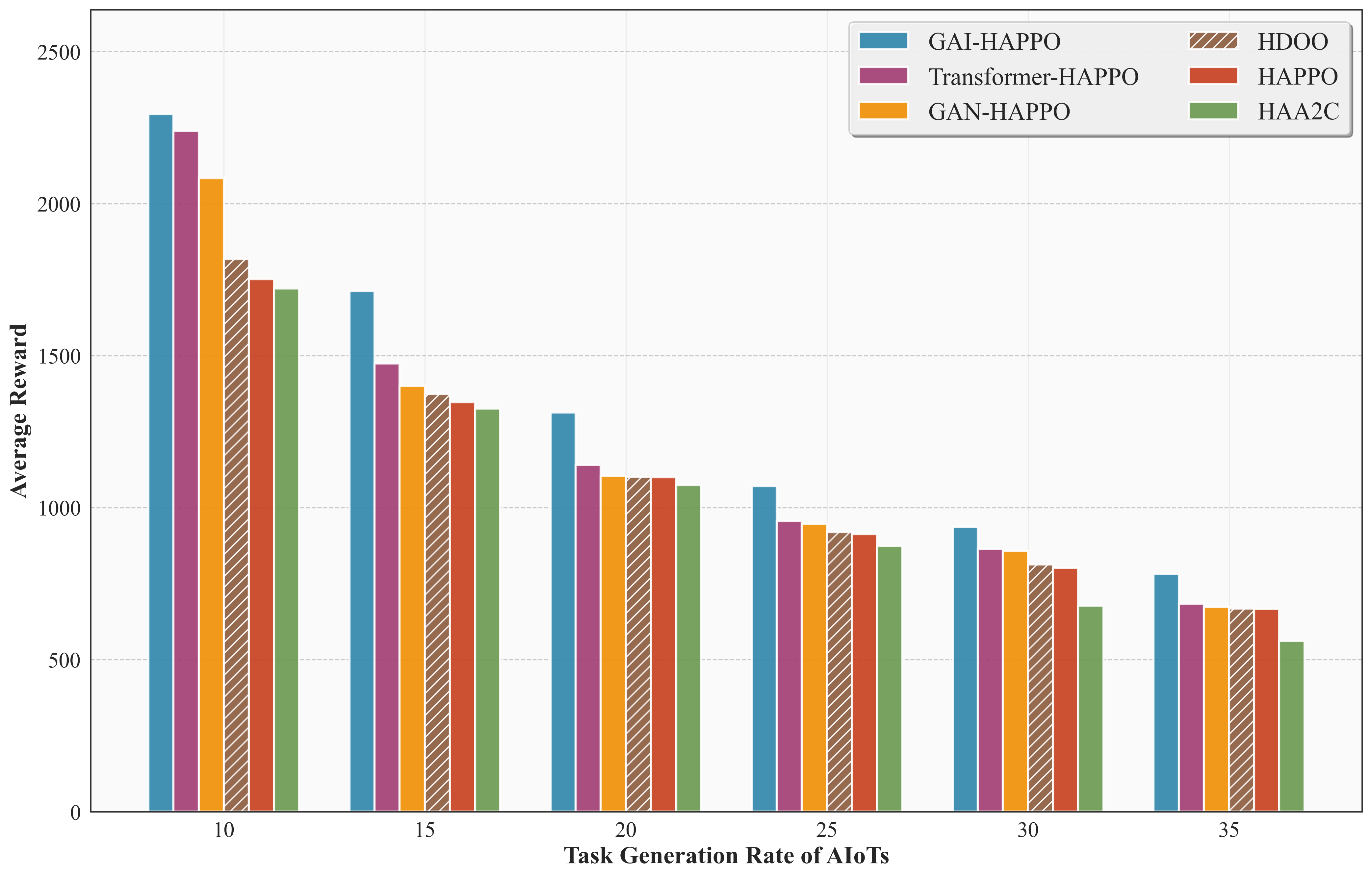}
}
\subfloat[Average execution time.\label{f9b}]{
    \includegraphics[width=5.8cm]{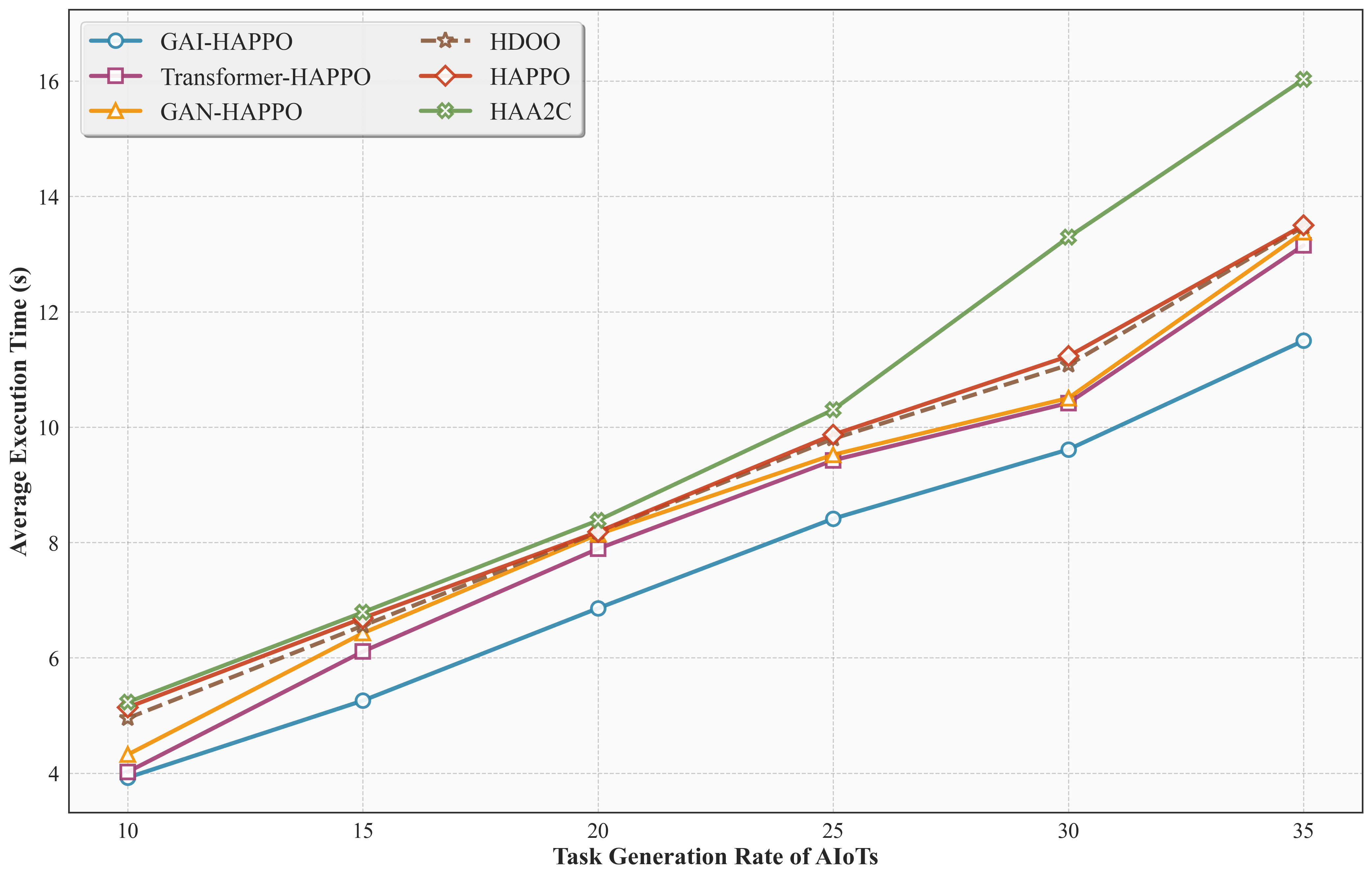}
}
\subfloat[Average computation of ratio.\label{f9c}]{
    \includegraphics[width=5.8cm]{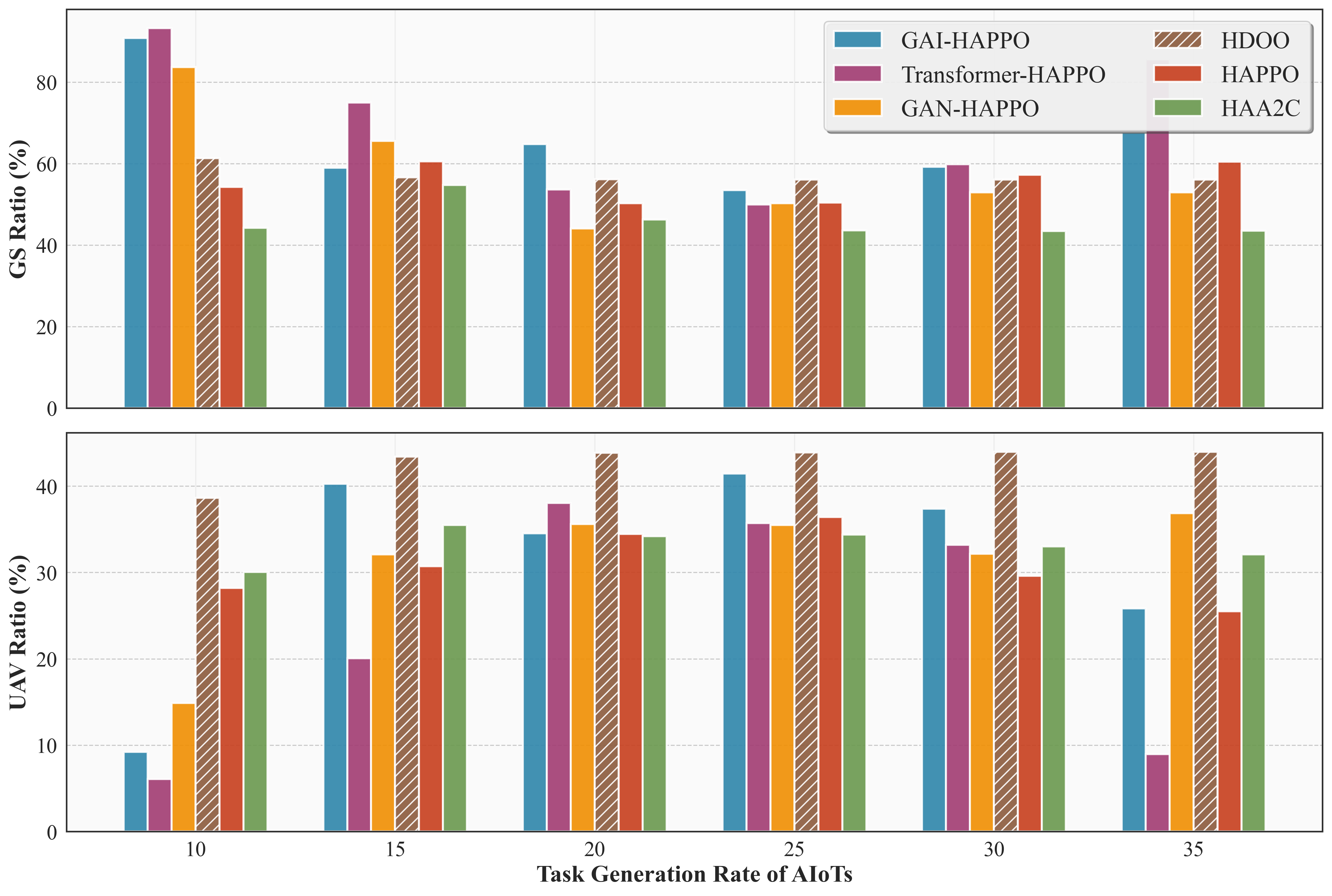}
}
\caption{Performance with different task generation rate of AIoTs.}
\label{f9}
\end{figure*}

\subsection{Evaluation on the Training of GAI-HAPPO and Benchmark Algorithms}
As shown in Fig.~\ref{f6}, all algorithms are trained for 2,500 episodes, with each episode containing 4,000 time steps. GAI-HAPPO consistently outperforms other methods in terms of average reward and convergence stability throughout training. Its learning curve steadily rises and converges smoothly, indicating both effective policy optimization and training robustness. Transformer-HAPPO achieves a comparable final reward but exhibits higher volatility, reflecting improved representational capacity but reduced training stability. 
Although GAN-HAPPO enhances the critic robustness through adversarial learning and achieves better results than HAPPO, its performance remains inferior to that of the full GAI-HAPPO integration. 
HAPPO and HAA2C lag behind in both reward and stability, with HAA2C demonstrating the weakest convergence. These trends highlight the advantage of jointly leveraging a transformer-based actor and GAN-based critic.

\subsection{Performance with Different Numbers of AIoTs}

Fig. \ref{f7}(a) depicts the average reward for each algorithm. GAI-HAPPO achieves the highest rewards across all AIoT numbers, demonstrating its effectiveness in decision optimization and resource coordination. Transformer-HAPPO follows closely but its performance degrades slightly as the AIoT count increases. GAN-HAPPO shows moderate rewards, benefiting from critic enhancement but lacking full coordination modeling. HDOO performs better than HAPPO and HAA2C under small-scale settings but fails to maintain competitive rewards as the number of AIoTs grows due to its static nature and lack of learning-based adaptation. HAPPO and HAA2C yield the lowest rewards, struggling to scale to more complex agent dynamics.

Fig. \ref{f7}(b) shows the average execution delay as AIoT numbers increase. GAI-HAPPO consistently outperforms others with minimal latency, particularly when the number of AIoTs is small to moderate. Transformer-HAPPO remains close but suffers slightly more delay in larger-scale scenarios. GAN-HAPPO delivers modest improvements, while HDOO, despite showing decent latency under low load, exhibits increasing delay as AIoT density increases, reflecting its inability to dynamically adapt. HAPPO and HAA2C result in the highest delays across the board.

Fig. \ref{f7}(c) evaluates the GS and UAV computation ratios. GAI-HAPPO and Transformer-HAPPO maintain balanced, scalable task allocation across nodes, demonstrating effective offloading strategies. GAN-HAPPO also shows balanced behavior but with less efficiency. HDOO exhibits a strong preference for UAVs, leading to UAV overuse and GS underutilization, which eventually contributes to execution delays and reward degradation. HAPPO and HAA2C show uneven distributions, leading to poor system utilization.

Overall, Fig. \ref{f7} highlights GAI-HAPPO as the most efficient algorithm, excelling in execution time, reward maximization, and scalability. Transformer-HAPPO remains competitive, while GAN-HAPPO achieves moderate performance. HDOO serves as a viable traditional baseline in sparse settings, but its static nature limits the scalability. In contrast, HAPPO and HAA2C are less suitable under high-demand conditions due to the limited adaptability.

\subsection{Performance with Different Numbers of UAVs}
Fig. \ref{f8}(a) shows the average reward achieved by each algorithm under varying numbers of UAVs. GAI-HAPPO consistently outperforms all baselines, demonstrating superior adaptability and coordination. Transformer-HAPPO follows closely, benefiting from structured representation learning. GAN-HAPPO achieves moderate rewards by improving critic robustness. HDOO performs better than HAPPO and HAA2C in low-to-moderate UAV settings due to its heuristic delay minimization but fails to scale with more UAVs, lacking adaptive coordinations. HAPPO and HAA2C yield the lowest rewards, reflecting poor generalization under dynamic UAV resources.

Fig. \ref{f8}(b) presents the average execution time across different UAV settings. GAI-HAPPO consistently achieves the lowest delay, highlighting its efficient policy learning. Transformer-HAPPO shows slightly higher latency but retains competitive performance. GAN-HAPPO remains moderate across all settings. HDOO exhibits relatively low delay when UAVs are few but becomes less effective as the number increases due to its static decision policy. HAPPO and HAA2C produce the highest delays, especially with larger UAV counts, reflecting inefficiencies in offloading and trajectory coordinations.

Fig. \ref{f8}(c) illustrates the UAV and GS computation ratios. GAI-HAPPO and Transformer-HAPPO maintain balanced task allocation across resources, reflecting effective coordination. GAN-HAPPO shows reasonable load distribution but slightly lower efficiency. HDOO exhibits over-reliance on UAVs, particularly under high-density conditions, leading to potential bottlenecks. HAPPO and HAA2C demonstrate imbalanced usage with underutilization of GSs, contributing to suboptimal performance.

Fig. \ref{f8} confirms that GAI-HAPPO offers the best performance across execution time, reward, and resource distribution. Transformer-HAPPO provides a strong alternative with slightly reduced scalability. GAN-HAPPO performs moderately well. HDOO serves as a practical heuristic in simpler scenarios but lacks adaptability. HAPPO and HAA2C consistently underperform due to limited modeling capabilities.

\subsection{Performance with Different Task Generation Rates of AIoTs}
Fig. \ref{f9}(a) shows the average reward for each algorithm under varying task generation rates. GAI-HAPPO consistently achieves the highest rewards, demonstrating strong adaptability in high-load scenarios. Transformer-HAPPO performs closely but declines under heavier loads. GAN-HAPPO maintains moderate rewards, while HDOO performs better than HAPPO and HAA2C at lower rates but degrades quickly under higher task rates. HAPPO and HAA2C achieve the lowest rewards overall, highlighting their limited optimization under dynamic workloads.

Fig. \ref{f9}(b) presents the average execution time. GAI-HAPPO maintains the lowest delays across all task rates, reflecting efficient policy learning and resource coordination. Transformer-HAPPO and GAN-HAPPO show moderate delays that increase with task rate. HDOO maintains low delay under light loads but performs poorly as demand increases due to lack of adaptability. HAPPO and HAA2C suffer the highest delays, with HAA2C showing the worst scalability.

Fig. \ref{f9}(c) illustrates the UAV and GS computation ratios. GAI-HAPPO and Transformer-HAPPO maintain well-balanced and stable task distribution. GAN-HAPPO shows reasonable balance, while HDOO increasingly shifts computation to UAVs at high rates, risking overload. HAPPO and HAA2C exhibit imbalanced allocation, further degrading performance.

Overall, GAI-HAPPO demonstrates the highest resilience and scalability under growing task loads, followed by Transformer-HAPPO and GAN-HAPPO. HDOO provides competitive performance at low-to-moderate rates but lacks robustness under pressure. HAPPO and HAA2C remain unsuitable for high-throughput applications.

\section{CONCLUSIONS}\label{s6}
In this paper, we design a cooperative UAV-GS-based MEC model to meet the computational and coverage demands of AIoTs in complex applications, such as environmental monitoring and maritime search and rescue.
The framework efficiently supports AIoTs in completing computational tasks by combining the mobility and flexibility of UAVs with the robust computational power of GSs.
To address the challenges of task uncertainties, AIoT trajectory unpredictability, device heterogeneity, and limited computational resources, we formulate a joint optimization problem involving task offloading and UAV trajectory planning, aimed at minimizing total execution time. 
In addition, we develop the GAI-HAPPO algorithm. GAI-HAPPO incorporates GAI models to improve the actor network's ability to model complex environments, extract high-level features, and adapt to dynamic conditions, while also stabilizing the critic network to mitigate MARL instability.
Extensive simulations show that our proposed GAI-HAPPO algorithm outperforms the original HAPPO baseline by 22.8\%, demonstrating its superior effectiveness in optimizing collaboration in heterogeneous MEC scenarios.
Future work will explore incorporating priority classes, enabling explicit task prioritization and ensuring critical tasks take precedence in multi-task scenarios.
\bibliography{references}

\end{document}